\documentclass[11 pt]{article}
\renewcommand{\baselinestretch}{1.47}
\date{}

\usepackage{graphicx,subfigure}
\usepackage{amsmath,amssymb}
\usepackage{cite}
\usepackage{url}
\usepackage{mystyle}

\allowdisplaybreaks
\begin{document}

\title{Optimal Joint Target Detection and Parameter Estimation By MIMO Radar}


\author{Ali Tajer\footnote{Electrical Engineering Department, Columbia
University, New York, NY 10027 (email: \{tajer, guido, wangx\}@ee.columbia.edu).}\and Guido H. Jajamovich\footnotemark[1]\and Xiaodong Wang\footnotemark[1]\and George V. Moustakides\footnote{Electrical Engineering Department, University of Patras, 26500 Rion, Greece (email: moustaki@ece.upatras.gr).}}

\maketitle

\begin{abstract}
We consider multiple-input multiple-output (MIMO) radar systems with widely-spaced antennas. Such antenna configuration facilitates capturing the inherent diversity gain due to independent signal dispersion by the target scatterers. We consider a new MIMO radar framework for detecting a target that lies in an {\em unknown} location. This is in contrast with conventional MIMO radars which break  the space into small cells and aim at detecting the presence of a  target in a specified cell. We treat this problem through offering a novel composite hypothesis testing framework for target detection when (i) one or more parameters of the target are unknown and we are interested in estimating them, and (ii) only a {\em finite} number of observations are available. The test offered optimizes a metric which accounts for both detection and estimation accuracies. In this paper as the parameter of interest we focus on the vector of time-delays that the waveforms undergo from being emitted by the transmit antennas until being observed by the receive antennas. The analytical and empirical results establish that for the proposed joint target detection and time-delay estimation framework, MIMO radars exhibit significant gains over phased-array radars for extended targets which consist of multiple independent scatterers. For point targets modeled as single scatterers, however, the detection/estimation accuracies of MIMO and phased-array radars for this specific setup (joint target  detection and time-delay estimation) are comparable.
\end{abstract}


\section{Introduction}
\label{sec:intro}
Inspired by the diversity gains yielded by multiple-input
multiple-output (MIMO) communication systems for alleviating
wireless fading impediments, the concept of MIMO radar systems has
been first introduced in \cite{Fishler:RC04} and further developed
in \cite{Fishler:Asilomar04, Fishler:SP06, Haimovich:SPM08,
Lehmann:Asilomar06, Stoica:SP07, Li:SPL07,Godrich:IT08_submitted}.
The underlying idea of MIMO radar for offering diversity gains is
to illuminate uncorrelated waveforms at the target such that their
corresponding reflections remain uncorrelated and hence offer
independent observations of the target. It has been demonstrated
that such diversities in observing the target bring about
performance improvement by enhancing the accuracy in detecting the target and estimating its associated parameters, e.g., range, speed, and direction.

MIMO radar systems can be exploited for capturing radar cross section (RCS) diversity via deploying widely-separated antennas \cite{Haimovich:SPM08}, or for establishing more degrees of freedom by configuring co-located antennas \cite{Li:SPM07}. While the former configuration improves detection and estimation qualities and supports high resolution target localization \cite{Lehmann:Asilomar06}, the latter one enhances the power of probing in the vicinity of target location \cite{Stoica:SP07, Li:SPL07, Bekkerman:SP06}.

In this paper we consider a widely-spaced antenna configuration and
treat two problems. First, we analyze the problem of target detection when some radar parameters are unknown and needed to be estimated. We offer a framework for joint target detection and parameter estimation (with optimality properties provided in Section \ref{sec:statement}) when  the receive antennas can acquire only a \emph{finite} number of observations. While the proposed framework can be exploited for detecting the target in conjunction with estimating any parameter of interest, we consider the problem of detecting a target that lies in an unknown location. In our formulation, the uncertainty about the target location is accounted for through the time-delays that the transmitted waveforms undergo before reaching the receive antennas. We formulate this problem as a composite hypothesis test which is shown to be solved optimally via the widely-known generalized likelihood ratio test (GLRT). Note that the existing optimality results of the GLRT hold only asymptotically for an \emph{infinite} number of observations under certain assumptions \cite[Sec. 5.6]{Levy:book}.

As the second problem, we analyze the diversity gain of the proposed
detector, which is defined as the decay rate of the mis-detection
probability with the increasing signal-to-noise ratio ($\snr$) for a
fixed false-alarm probability~\cite{Haimovich:SPM08}. We
analytically quantify the diversity gain as a function of the number
of transmit and receive antennas. This notion of diversity gain for
MIMO radars has been first examined for MIMO radars in
\cite{Fishler:SP06} and  \cite{Haimovich:SPM08} for the MIMO radar
model introduced in \cite{Fishler:SP06}, which considers detecting
the presence of a target at a given location. It is noteworthy that
another notion of diversity gain defined as the detector's
divergence in high $\snr$ regimes has also been analyzed in [3].

We treat the two aforementioned problems for two target models: point targets which are modeled as single scatterers \cite{Haimovich:SPM08, Lehmann:Asilomar06}; and extended targets, which are comprised of many  isotropic and independent scatterers. The summary of the contributions and results of this work is as follows.

We propose an optimality measure which is shaped by target detection
performance, parameter estimation accuracy and false-alarm
probability. The corresponding optimal composite hypothesis test
that satisfies some optimality criteria is introduced and deployed
for detecting a target in an unknown location. Next, for an $N_t\times N_r$ MIMO radar system we characterize the maximum likelihood (ML) estimate of the time-delay vector which consists of $N_tN_r$ components corresponding to different pairs of transmit-receive antennas. The optimal detector corresponding to such ML estimate is also derived. This detector differs from the existing ones in the literature in the sense that it aims at detecting a target in an unknown location, whereas the existing MIMO radar detectors break the space into small cells and detect the presence of the target in a specific cell. Finally, we demonstrate that for the proposed detector, in an $N_t\times N_r$ MIMO radar system with widely-spaced antennas and with an upper-limit constraint on the false-alarm probability, the  mis-detection probability decays as $\snr^{-N_tN_r}$ for extended targets and decays as $\snr^{-1}$ for point targets. Such mis-detection probability in phased-array radars for both extended and point
targets is shown to decay as $\snr^{-1}$. The same observations for
extended targets and phased-array radars have also been made in
\cite{Fishler:SP06} and \cite{Haimovich:SPM08}, albeit for a
different MIMO radar model and different detectors.

The remainder of the paper is organized as follows. Section~\ref{sec:model} provides the MIMO radar system model. The
statement of the composite detection problem and the definition of
optimality are provided in Section~\ref{sec:statement}. We formulate and analyze the joint estimation and detection problem for MIMO radars for extended and point targets in  Sections~\ref{sec:extended}~and~\ref{sec:point}, respectively, and for phased-array radars in Section~\ref{sec:PA}. The analysis on the diversity gain are presented in Section~\ref{sec:diversity}. Simulation results are illustrated in Section~\ref{sec:simulations} and Section~\ref{sec:conclusions} provides the concluding remarks.

\section{MIMO Radar}
\label{sec:model}
We consider a MIMO radar system comprising of $N_t$ and $N_r$ transmit and receive antennas, respectively, and adopt the classical Swerling case I model \cite{Skolnik:book} extended for multiple-antenna systems \cite{Fishler:SP06, Haimovich:SPM08}. According to this model, a target consists of one or more small scatterers exhibiting random, independent and isotropic scintillation.

We define $P$ as the number of the target's scatterers and denote the locations of these scatterers in the Cartesian coordination by $X_p$ for $p=1,\dots,P$. Also, we denote the reflectivity factor of the $p^{th}$ scatterer by $\zeta_p$ and assume that $\{\zeta_p\}_{p=1}^P$ are identically and independently distributed as zero-mean complex random variables with variance $\frac{1}{P}$. The target and reflectivity factors are assumed to remain constant during a finite number of observations denoted by $K$ and change to independent states afterwards.
\begin{figure}[t]
  \centering
  \begin{picture}(20,20)
    \put(0,158){$X^t_1$}
    \put(0,109){$X^t_m$}
    \put(0,60){$X^t_{N_t}$}
    \put(80,-15){$X^r_1$}
    \put(130,-15){$X^r_n$}
    \put(180,-15){$X^r_{N_r}$}
    \put(280,195){$X_p$}
    \put(230,230){{Target}}
    \put(130,155){$d'$}
    \put(175,110){$d''$}
    \put(210,110){$d^p_{m,n}=d'+d''$}
  \end{picture}
  \includegraphics[width=3.5 in]{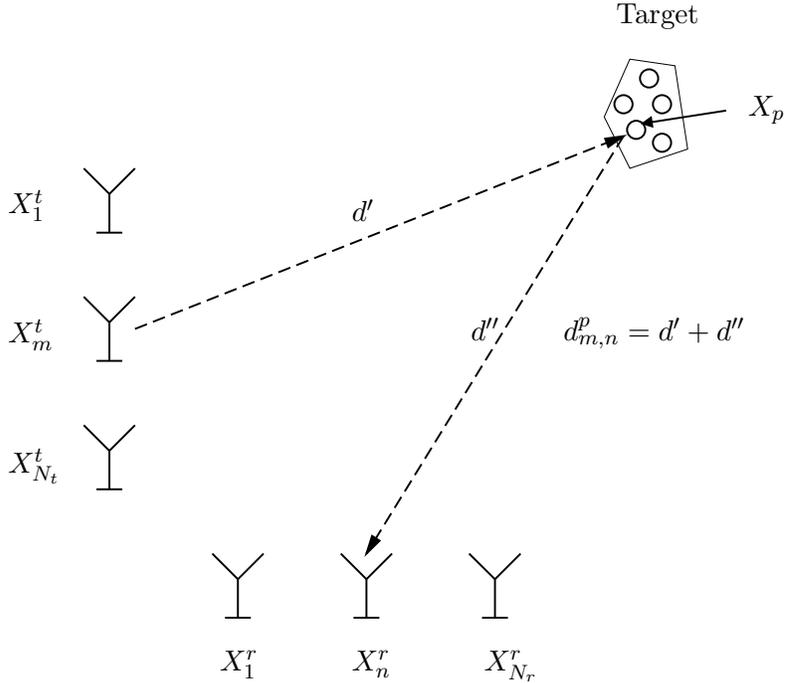}\\
  \caption{A MIMO radar system.}\label{fig:radar}
\end{figure}
Motivated by capturing the inherent diversity provided by
independent scatterers, the antennas are spaced widely enough (such
that they satisfy the conditions in \cite[Sec. II.A]{Fishler:SP06})
to ensure that the received antennas capture uncorrelated
reflections from the target. As illustrated in Fig. \ref{fig:radar}, we assume that the transmit antennas are located at $X^{t}_m$, for $m=1,\dots,N_t$, and the receive antennas are located at $X^{r}_n$, for $n=1,\dots,N_r$. The transmit antennas emit $N_t$ \emph{narrowband} waveforms of duration ${T}$ with baseband equivalent models given by $\sqrt{\frac{E}{N_t}}s_m(t)$ for $m=1,\dots,N_t$, where $E$ is the total transmitted energy and $\int_{{T}}|s_m(t)|^2dt=1$. In contrast to the conventional phased-array radars which deploy waveforms that are identical upto a scaling factor \cite{Haykin:book}, in MIMO radar systems these waveforms are designed such that they facilitate acquiring independent observations of each scatterer and often are assumed to be orthonormal \cite{Haimovich:SPM08}, i.e.,
\begin{align}\label{eq:orthonormal}
    \forall m,n,\quad \int_{T} s_m(t)s^*_n(t)\;dt=\delta(m-n),
\end{align}
where $^*$ denotes complex conjugate and $\delta(\cdot)$ is the
Dirac's delta function. As depicted in Fig. \ref{fig:radar}, the
waveform illuminated by the $m^{th}$ transmit antenna to the
$p^{th}$ scatterer and received by the $n^{th}$ receive antenna
passes through an end-to-end distance which we denote by $d^p_{m,n}$
and undergoes a time-delay which we denote by $\tau^p_{m,n}\dff
d^p_{m,n}/c$, where $c$ is the speed of light. By defining $\beta$
as the path-loss exponent and superimposing the effects of all
scatterers, the base-band equivalent of the signal received by the
$n^{th}$ receive antenna due to the waveform $s_m(t)$ (transmitted by
the $m^{th}$ transmit antenna) is given by
\begin{equation}\label{eq:model1}
    r_{m,n}(t)=\sqrt{\frac{E}{N_t}}\sum_{p=1}^P\zeta_p\Big(\frac{1}{d^p_{m,n}}\Big)^\beta s_m(t-\tau_{m,n}^p)\;e^{-j2\pi f_c\tau^p_{m,n}}+z_{m,n}(t).
\end{equation}
Note that this model differs from those of \cite{Fishler:SP06} and
\cite{Haimovich:SPM08} in the sense that we have added the attenuation
effects of path-losses by including the terms $\Big(\frac{1}{d^p_{m,n}}\Big)^\beta=\Big(\tau^p_{m,n}\;c\Big)^{-\beta}$.
The exponential term $\exp(-j2\pi
f_c\tau^p_{m,n})$ in (\ref{eq:model1}) represents the effect of propagation phase shift,
where $f_c$ is the carrier frequency, and $z_{m,n}(t) \sim
\mathcal{CN}(0,\frac{1}{N_t})$, denotes the additive white Gaussian noise.

We define $X_0$ as the location of the gravity center of the target and denote its associated time-delays and distances by $\{\tau_{m,n}\}$ and $\{d_{m,n}\}$, respectively. We assume that the distances $\{d_{m,n}\}$ are considerably larger than the dimensions of the object such that we can replace the distances and the time-delays associated with the scatterer $X_p$ with those corresponding to the gravity center of the target $X_0$, i.e., $d^p_{m,n}=d_{m,n}$ and $\tau^p_{m,n}=\tau_{m,n}, \forall p$. Therefore
\begin{equation}\label{eq:approximate}
    s_m(t-\tau^p_{m,n})= s_m(t-\tau_{m,n}), \quad\mbox{for}\;\;m=1,\dots,N_t, \;\;n=1,\dots,N_r,\;\;\mbox{and}\;\;p=1,\dots,P.
\end{equation}
Using (\ref{eq:model1})-(\ref{eq:approximate}), the received signal
at the $n^{th}$ antenna, which is a superposition of all emitted
waveforms, is given by
\begin{eqnarray}
    \nonumber r_n(t)&=&\sqrt{\frac{E}{N_t}}\;\sum_{m=1}^{N_t}\sum_{p=1}^P\Big(\frac{1}{d_{m,n}}\Big)^\beta\;\zeta_p s_m(t-\tau_{m,n})\;e^{-j2\pi f_c\tau^p_{m,n}}+z_{n}(t)\\
    \label{eq:model_cont}&=&\sqrt{\frac{E}{N_t}}\;\sum_{m=1}^{N_t}\frac{c^{-\beta}}{\tau^\beta_{m,n}}\;s_m(t-\tau_{m,n}) \underset{\dff\;h_{m,n}}{\underbrace{\sum_{p=1}^P\;\zeta_p \;e^{-j2\pi f_c\tau^p_{m,n}}}}+z_{n}(t),
\end{eqnarray}
where $z_n(t)\dff\sum_{m=1}^{N_t}z_{m,n}(t)\sim\mathcal{CN}(0,1)$. Furthermore, we assume that the waveforms are narrowband. Based on the narrow-band assumption we get \cite{Haimovich:SPM08}
\begin{equation}\label{eq:narrow}
    \forall \tau,\quad s_m(t)= e^{j2\pi f_c\tau}\;s_m(t-\tau), \;\;\mbox{for}\;\;m=1,\dots,N_t,
    \;\;\mbox{and}\;\;n=1,\dots,N_r,
\end{equation}
which in conjunction with the orthonormality assumption
(\ref{eq:orthonormal}) implies that
\begin{align}\label{eq:orthonormal2}
    \forall\;\tau_{m,k},\tau_{n,k},\;\;\;\int_T
    s_m(t-\tau_{m,k})s^*_n(t-\tau_{n,k})\;dt=\delta(m-n).
\end{align}
We also define the time-delay vector $\btau\dff[\tau_{1,1},\dots,\tau_{N_t,N_r}]$. Based on the model given in (\ref{eq:model1}) and noting that the noise-terms $\{z_n(t)\}$ are unit-variance, the transmission signal-to-noise ratio, denoted by $\snr$, is given by $\snr=\frac{E}{T}$.

\section{Problem Statement}
\label{sec:statement}
We assume that the receive antennas sample the received signal at the rate of $\frac{1}{T_s}$ samples per second. By defining $r_n[k]\dff r_n(kT_s)$, $z_{n}[k]\dff z_{n}(kT_s)$ and $s_m[k;\tau]\dff s_m(kT_s-\tau)$, the discrete-time baseband equivalent of the received signal when a target is present is given by
\begin{equation}\label{eq:model_discrete}
    r_n[k]=\sqrt{\frac{E}{N_t}}\;c^{-\beta}\;\sum_{m=1}^{N_t} \frac{1}{\tau^\beta_{m,n}}\;h_{m,n}s_m[k;\tau_{m,n}]+z_{n}[k], \;\quad\mbox{for}\quad k=1,\dots,K.
\end{equation}
We also assume that the sampling rate is high enough to ensure that the discrete-time signals $s_m[k;\tau_{m,n}]$ remain orthogonal for  arbitrary delays $\tau_{m,l}, \tau_{n,l}$, i.e.,  $\sum_ks_m[k;\tau_{m,l}]s_n^*[k;\tau_{n,l}]=\frac{1}{T_s}\delta(m-n)$.
Let us define $\br[k]\dff [r_1[k],\dots,r_{N_r}[k]]^T$  for
$k=1,\dots,K$ and $\bR\dff [\br[1]^T,\dots,\br[K]^T]^T$.

The target detection framework that we propose is different from the
conventional target detection with MIMO radars. In conventional MIMO
radars, e.g., \cite{Fishler:SP06} and references therein, the space
is broken into small cells and the radar detects the presence of the
target in a given cell. In such radar models, the location to be
probed is given, which means that the corresponding time-delay
vector $\btau$ is also given. Therefore, the detection problem can
be cast as testing whether a target exists at a given set of
time-delays (c.f. \cite{Fishler:SP06}). In contrast, our objective
is to detect the presence of a target that lies in an unknown
location and thereof has unknown corresponding time-delays. For this purpose we treat the time-delay vector $\btau$ as the unknown parameter to be estimated and based on that we formulate the target detection
problem. In other words we exploit $\btau$ as an intermediate
variable which we estimate for being able to carry out target
detection. Therefore, in our framework the main objective of
estimating the time-delays is to facilitate performing detection. As
a minor application, such time-delays can also be exploited for
estimating the location of the target. Such target localization, however, is not optimal and for achieving optimal target localization one needs to treat the target location as the unknown parameter of the
interest and deploy the same framework for carrying out joint target
detection and target localization (instead of time-delay
estimation).

Let $f_0(\bR)$ denote the probability density function (pdf) of
the received signal when a target is not present. When a target is
present, the pdf of the received signal depends on an unknown
parameter $\bx$ (in this paper $\bx$ is the vector of time-delays) and is denoted by $f_1(\bR\med \bx)$. Therefore, by defining the estimate of $\bx$ by $\hat\bx$, the detection part of the problem can be cast as
\begin{align}\label{eq:hypothesis1}
    \left\{\begin{array}{l}
      \mathcal{H}_0: \quad\quad \bR \sim f_0(\bR), \\
      \mathcal{H}_1: \quad\quad \bR \sim f_1(\bR\med \hat\bx).
    \end{array}\right.
\end{align}
Our objective is to jointly characterize the estimator $\hat\bx$ and
find the test for deciding between the hypotheses ${\cal H}_0$ and
${\cal H}_1$. For this purpose we define and optimize a measure
which combines estimation and detection accuracies. The underlying
idea for performing such optimization is to deploy the notions of
{\em randomized tests} and {\em randomized estimators}. For any
given observation vector $\bR\in\mathbb{C}^{|\bR|}$ we assign
complementary probability terms $\delta_0(\bR)$ and
$\delta_1(\bR)=1-\delta_0(\bR)$ to the hypotheses $\mathcal{H}_0$
and $\mathcal{H}_1$, respectively. In a randomized test,
$\delta_i(\bR)$ for $i=0, 1$ is the probability that hypothesis
$\mathcal{H}_i$ is selected. Clearly deterministic tests are special
cases of randomized tests where $\{\delta_i(\bR)\}$ take only the
values $0$ or $1$. Whenever the decision is in favor of
$\mathcal{H}_1$, we also have to estimate the unknown parameter
$\bx$. In deterministic approaches, the estimate of $\bx$ is a
deterministic function of the observed data. On the contrary,
randomized estimators, instead of generating a deterministic value
$\hat\bx$, generate a random variable $\hat\bx$ according to a
distribution which is a function of the observed data. Deterministic
estimators can be classified as a special case of randomized
estimators.

We strive to identify the decision rules $\delta_0(\bR)$ and $\delta_1(\bR)$ such that the error in estimating $\bx$, conditioned on the hypothesis $\mathcal{H}_1$ being true, is minimized. This minimization is also subject to an upper bound constraint on the false alarm probability $\pr(\mathcal{H}_1\med \mathcal{H}_0)$. When the hypothesis $\mathcal{H}_1$ is true, there exist two types of errors in estimating $\bx$. First, the test result can be correctly in favor of $\mathcal{H}_1$, but $\bx$ is estimated erroneously, where we denote the cost function associated with such estimation error by $C(\hat\bx,\bx)$. Secondly, the test output might be wrongly in favor of $\mathcal{H}_0$, which leads to mis-estimating $\bx$ and results in an error cost which we denote by $C_0(\bx)$. Note that the cost function $C(\hat\bx, \bx)$ can be selected arbitrarily. In this paper we consider the mean-square error (MSE) cost given by $C(\hat\bx,\bx)=\|\hat\bx-\bx\|^2$ and $C_0(\bx)=\|\bx\|^2$.

Subject to a maximum tolerable level of false-alarm probability $P_{fa}$, the estimation strategy which {\em minimizes the average cost} over all possible randomized estimators is given by
\begin{equation}\label{eq:P2}
    \mathcal{P}(\bR)=
    \left\{\begin{array}{ll}
      \min_{\{\delta_i(\bR)\}} &\bbe_{\bR,\bx,\hat\bx}\left[C(\hat{\bx},\bx)\med \mathcal{H}_1 \right] +\bbe_{\bR,\bx}\left[C_0(\bx)\med \mathcal{H}_1 \right]\\
      \mbox{s.t.}& \pr\left(\mathcal{H}_1\med
  \mathcal{H}_0\right)\leq P_{fa}.
    \end{array}\right.
\end{equation}
The optimal hypothesis test (formalized by $\{\delta_i(\bR)\}$) and the optimal estimation strategy yielded by solving the problem ${\cal P}(\bR)$ for the choices of $C(\hat\bx,\bx)=\|\hat\bx-\bx\|^2$ and $C_0(\bx)=\|\bx\|^2$ are described in the following theorem.
\begin{theorem}[Moustakides \cite{Moustakides:IT09_submitted}]\label{th:GLRT}
For a finite cardinality vector $\bR\in\mathbb{C}^{|\bR|}$ and an
unknown vector parameter $\bx\in{\cal A}$, solving ${\cal P}(\bR)$
provides the optimal estimation strategy $\hat\bx$ and the
subsequent optimal detection rule for deciding between ${\cal H}_0$
and ${\cal H}_1$ are given by
\begin{equation}\label{eq:GLRT2}
 \hat{\bx}=\arg\max_{\bx\in{\cal A}} \pi(\bx)f_{1}(\bR\med\bx),\quad \textrm{ and }\quad\quad \frac{f_{1}(\bR\med\hat{\bx})}{f_{0}(\bR)}
  \underset{\mathcal{H}_0}{\overset{\mathcal{H}_1}\gtrless}\lambda,
\end{equation}
where $\pi(\bx)$ is the prior distribution of $\bx$ and the estimation cost functions are $C(\hat\bx,\bx)=\|\hat\bx-\bx\|^2$ and $C_0(\bx)=\|\bx\|^2$.
\end{theorem}

The above theorem essentially establishes the GLRT as the optimal estimation/detection strategy with properties P1-P4 enumerated in the sequel. Several other {\em asymptotic} optimality results are known for the GLRT which are all based on having an {\em infinite} number of observation (c.f. \cite{Moustakides:IT09_submitted} and \cite[Chapter 22]{Kendall:book}). $\hat\bx$ given in (\ref{eq:GLRT2}) is the {\em maximum a posteriori} (MAP) estimate of $\bx$. As we do not have any prior information about the parameter to be estimated (the vector of time-delays $\btau$), throughout the  analysis we assume that $\pi(\bx)$ has a uniform distribution. Hence,  the MAP estimate of $\bx$ becomes its ML estimate. The estimator and detector provided by Theorem \ref{th:GLRT} have the following properties.
\begin{description}
  \item [\rm P1)] The ML estimator $\hat\bx$ minimizes the {\em average} minimum-mean square estimation error of $\bx$.
  \item [\rm P2)] The false-alarm probability of the target detector is kept below a certain level.
  \item [\rm P3)] For the given set of ML estimates $\hat\bx$, the target detector is Bayesian-optimal, i.e., the Bayes risk is minimized \cite[Sec. II.B]{Poor:book}.
  \item [\rm P4)] The test requires only a finite number of samples, i.e., $|\bR|<\infty$.
\end{description}
In order to apply the theorem above to our joint estimation/detection problem we set the vector of time-delays as the parameter to be  estimated, i.e., $\bx=\btau$. For deploying the optimal estimator and  detector provided in (\ref{eq:GLRT2}) we also need to characterize the domain of $\bx=\btau$ which in Theorem \ref{th:GLRT} is denoted by ${\cal A}$. Note that the vector $\btau$ cannot be any arbitrary vector in the space  $\mathbb{R}_+^{N_tN_r}$. This is due to the fact that for any given  placement of antennas, there exists a correlation between the elements of $\btau$ and there are some equations relating these time-delays. For the given antenna placements $\{X^t_m\}_{m=1}^{N_t}$ and $\{X^r_n\}_{n=1}^{N_r}$ let us define ${\cal   X}\dff\{X^t_m,X^r_n\}$. Also for the given antenna placement ${\cal X}$ define ${\cal A}({\cal  X})\subseteq\mathbb{R}_+^{N_tN_r}$ as the set which contains the vectors that can be valid choices for $\btau$.

Precise characterization of ${\cal A}({\cal X})$ is crucial for attaining the optimal estimation and detection performance. Any inaccuracy in such characterization degrades the performance and provides only {\em sub-optimal} estimators and detectors. In the sequel we provide necessary conditions for a given $\btau$ to be a valid  time-delay vector. Such necessary conditions describe a set of time-delay vectors which is a superset of ${\cal A}({\cal X})$. We also {\em conjecture} that these conditions are sufficient which consequently can characterize ${\cal A}({\cal X})$ precisely. If these conditions are not sufficient, then the resulting approximation of ${\cal A}({\cal X})$ will only provide a {\em sub-optimal} solution.

For establishing the necessary conditions let us break down the
end-to-end time-delay $\tau_{m,n}$ as $\tau_{m,n}=t_m+t'_n$ where
$t_m$ is the time required for the waveform to travel from the
$m^{th}$ transmit antenna to the target and $t'_n$ is the time
required for the reflected waveform to travel from the target to the
$n^{th}$ receive antenna. If a given $\btau$ is a valid time-delay
vector then $\forall m,i\in\{1,\dots,N_t\}$ the three terms $t_m$,
$t_i$, and $\frac{\|X^t_m-X^t_i\|}{c}$ should constitute the lengths
of a triangle with its vertices given by the locations of the
$m^{th}$ transmit antenna, the $i^{th}$ transmit antenna, and the
target ($\|X^t_m-X^t_i\|$ is the distance between the $m^{th}$ and
$i^{th}$ transmit antennas). The triangular inequalities impose that
$\btau$ is a valid time-delay vector only if
\begin{equation*}
    |t_m-t_i|\leq \frac{\|X^t_m-X^t_i\|}{c}\quad\mbox{for}\quad m,i=1,\dots,N_t.
\end{equation*}
Similarly, by considering the triangles corresponding to the receive antennas we get the necessary conditions
\begin{equation*}
    |t'_n-t'_j|\leq \frac{\|X^r_n-X^r_j\|}{c}\quad\mbox{for}\quad n,j=1,\dots,N_r.
\end{equation*}
By recalling that $\tau_{m,n}=t_m+t'_n$ $\;\forall\;m,n,i,j$, we find
that $|t_m-t_i|=|\tau_{m,n}-\tau_{i,n}|$ and
$|t'_n-t'_j|=|\tau_{m,n}-\tau_{m,j}|$. Therefore, for the subspace ${\cal A}({\cal X})$ we have
\begin{equation}\label{eq:A}
    {\cal A}({\cal X})\subseteq\left\{\btau:|\tau_{m,n}-\tau_{i,n}|\leq\frac{\|X^t_m-X^t_i\|}{c}\ ,\;\;\mbox{and}\;\; |\tau_{m,n}-\tau_{m,j}|\leq\frac{\|X^r_n-X^r_j\|}{c} \;\;\forall\; m,i,j,n \right\}.
\end{equation}
We conjecture that any vector $\btau$ that satisfies the conditions above is a valid time-delay vector and therefor the subset operator $\subseteq$ can be replaced by equality. We underline that, however, in the case that these conditions are not sufficient, performing optimization on the above superset of ${\cal A}({\cal X})$ will provide only {\em sub-optimal} estimators and detectors.

\section{Joint Detection and Estimation for Extended Targets}
\label{sec:extended}

In this section we consider the targets that are extended enough to be
modeled as a group of isotropic and independent scatterers, i.e.,
$P\gg 1$ in (\ref{eq:model1}). Our objective is to optimally  detect a target, based on the model in (\ref{eq:hypothesis1}), and simultaneously estimate the vector of time-delays $\btau$. At the same time we assume that only a finite number of observations are available. According to Theorem \ref{th:GLRT} this problem can be solved via the GLRT given in (\ref{eq:GLRT2}) for $\bx=\btau$ and for ${\cal A}={\cal A}({\cal X})$ given in (\ref{eq:A}).

We start by deriving the ML estimate of $\btau$ and then provide the detection-related analysis. We also develop a technique in Section \ref{sec:simulations} based on the Newton-Raphson method for localizing the target by using the estimated time-delay vector. Exploiting this simple target localization method is not the optimal strategy and is merely intended to provide some numerical evaluations to show the improvement gained by deploying MIMO radars.

\subsection{Time Delay Estimation}
\label{sec:multi_estimation}
The hypothesis given in (\ref{eq:hypothesis1}) can be cast as
\begin{align}\label{eq:multi_hypothesisET}
    \left\{\begin{array}{ll}
      \mathcal{H}_0: & r_n[k]=z_{n}[k],\\
      \mathcal{H}_1: &     r_n[k]=\sqrt{\frac{E}{N_t}}\;c^{-\beta}\sum_{m=1}^{N_t}\frac{1}{\tau_{m,n}^\beta}\;h_{m,n}\;
    s_m[k;\tau_{m,n}]+z_{n}[k].
    \end{array}\right.
\end{align}
Define $\bh\dff[h_{1,1},h_{1,2},\dots,h_{N_t,N_r}]^T$, where
$h_{m,n}$, as defined in (\ref{eq:model_cont}), accounts for the
effects of the position and reflectiveness of the scatterers
corresponding to the $m^{th}$ transmit and the $n^{th}$ receive
antennas. The antennas are widely separated and the reflectivity
factors $\{\zeta_p\}$ are complex and independently distributed with
zero mean and variance $\frac{1}{P}$ with $P\gg 1$. Therefore, by
using the central limit theorem, $\{h_{m,n}\}$ are i.i.d. and
distributed as $\mathcal{CN}(0,1)$ \cite{Fishler:SP06}.

Based on the model given in (\ref{eq:model_discrete}), when a target is present, the vector of the received signals $\bR$ depends on the time-delays, which we are interested in estimating, as well as on the unknown random vector $\bh$. Therefore, when a target is present, the pdf of $\bR$ for any given $\btau$ and $\bh$ is $f_1(\bR\med\btau,\bh)$. In order to obtain the ML estimate of $\btau$ one approach is to estimate it through solving $\arg\max_{\btau\in{\cal A}({\cal X})}f_1(\bR\med\btau)$, which requires recovering $f_1(\bR\med\btau)$ from  $f_1(\bR\med\btau,\bh)$ by averaging over all realizations of $\bh$. Alternatively, we can estimate it jointly with $\bh$ when doing so is deemed to be beneficial. The latter approach is more beneficial when an accurate estimate of $\bh$ is available, e.g., in the high $\snr$ regimes, while averaging leads to a better performance when the estimate is very inaccurate, e.g., in the low $\snr$ regimes. The ML estimate of $\btau$ is provided in the following theorem for both scenarios.

\begin{theorem}\label{th:multi_ML}
The ML estimate of $\btau$ for extended targets and a given antenna placement ${\cal X}$
\begin{enumerate}
  \item through MAP estimation of $\bh$ is given by
  \begin{align*}
      \hat\btau^{\rm MAP}=
      \arg\max_{\btau\in{\cal A}({\cal X})}\sum_{m=1}^{N_t}\sum_{n=1}^{N_r}\frac{\left|\sum_{k=1}^{K}r_n[k]\;s_m^*[k;\tau_{m,n}]\right|^2} {\frac{1}{T_s}+\frac{N_t}{E}(c\tau_{m,n})^{2\beta}},
\end{align*}
  \item and through averaging over all realizations of $\bh$ is given by
  \begin{equation*}
    \hat\btau^{\rm ave}=
      \arg\max_{\btau\in{\cal A}({\cal X})}\sum_{m=1}^{N_t}\sum_{n=1}^{N_r}\left\{\frac{\left|\sum_{k=1}^{K}r_n[k]\;s_m^*[k;\tau_{m,n}]\right|^2} {\frac{1}{T_s}+\frac{N_t}{E}(c\tau_{m,n})^{2\beta}}- \log\left(\frac{E}{T_sN_t}(c\tau_{m,n})^{-2\beta}+1\right)\right\},
  \end{equation*}
\end{enumerate}
\end{theorem}
\begin{proof}
See Appendix \ref{app:th:multi_ML}.
\end{proof}


\subsection{Target Detection}
\label{sec:multi_detection}
Based on the ML estimates of the time-delay vector $\btau$ provided
in Theorem \ref{th:multi_ML}, we proceed to find the corresponding
optimum detectors. We show that both estimates give rise to the same
optimal detector given in the following lemma.
\begin{lemma}
\label{lemma:MIMO_Extended_Det}
The optimal test for extended targets and for the given estimate $\hat\btau$ is
\begin{eqnarray}
 \label{eq:multi_test}
 \sum_{m=1}^{N_t}\sum_{n=1}^{N_r}\frac{\bigg| \sum_{k=1}^{K}r^*_n[k]\;s_m[k;\hat\tau_{m,n}]\bigg|^2}{\frac{E}{T_sN_t}+(c\hat\tau_{m,n})^{2\beta}} \underset{\mathcal{H}_0}{\overset{{\cal H}_1}{\gtrless}}\theta.
\end{eqnarray}
\end{lemma}
\begin{proof}
Based on the model in (\ref{eq:multi_hypothesisET}) the likelihood $f_0(\bR)$ under the null hypothesis is given by
\begin{eqnarray}
  \label{eq:multi_pdf_H0} f_0(\bR)&=&(\pi)^{-KN_r}
  \exp\left({-\sum_k\|\br[k]\|^2}\right).
\end{eqnarray}
Finding the estimate $\hat\btau^{\rm MAP}$ is carried out by jointly estimating $\btau$ and $\bh$. For this estimate of $\btau$, by setting $\bx=[\btau,\;\bh]$ the optimal detector characterized in (\ref{eq:GLRT2}) is found as
\begin{eqnarray*}
 \frac{\max_{\bx\in{\cal A}} \pi(\bx)f_{1}(\bR\med\bx)}{f_{0}(\bR)}=\frac{\max_{\btau\in{\cal A}({\cal X}),\bh} f_{1}(\bR\med\btau,\bh)f_{\bh}(\bh)}{f_{0}(\bR)}=\frac{f_{1}(\bR\med\hat\btau,\hat\bh)f_{\bh}(\hat\bh)}{f_{0}(\bR)}
  \underset{\mathcal{H}_0}{\overset{\mathcal{H}_1}\gtrless}\lambda,\quad
\end{eqnarray*}
where $\hat\btau$ and $\hat\bh$ are the MAP estimates obtained in Theorem \ref{th:multi_ML}. By following the same line of argument as in Appendix \ref{app:th:multi_ML}  (\ref{eq:multi_pdf_H1_1})-(\ref{eq:multi_pdf_H1_5}) and recalling the definitions of $\ba_{\btau}$ and $\bB_{\btau}$ given in (\ref{eq:multi_pdf_H1_3}) and (\ref{eq:B}), respectively, the detector is given by
\begin{eqnarray}
 \nonumber
  \log\left(\frac{f_1(\bR\med\boldsymbol{\hat\tau},\hat{\bh})f_{\bh}(\hat\bh)}{f_0(\bR)}\right)&=&
  \log\left(\frac{(\pi)^{-KN_r}\exp\left({-\sum_k\|\br[k]\|^2}\right) \exp\left(\ba_{\hat\btau}^H\bB^{-1}_{\hat\btau}\ba_{\hat\btau}\right)}{(\pi)^{-KN_r}
  \exp\left({-\sum_k\|\br[k]\|^2}\right)}\right)\\
  &=&  \ba_{\hat\btau}^H\bB^{-1}_{\hat\btau}\ba_{\hat\btau}\\
  &=&
 \sum_{m=1}^{N_t}\sum_{n=1}^{N_r}\frac{\bigg| \sum_{k=1}^{K}r^*_n[k]\;s_m[k;\hat\tau_{m,n}]\bigg|^2}{\frac{E}{T_sN_t}+(c\hat\tau_{m,n})^{2\beta}} \underset{\mathcal{H}_0}{\overset{{\cal H}_1}{\gtrless}}\theta,
\end{eqnarray}
where $\theta\dff \frac{2N_t}{E}\log\lambda$. We further define $b_{m,n}\dff \sum_{k=1}^{K}r^*_n[k]\;s_m[k;\hat\tau_{m,n}]$ and $l_{m,n} \dff \frac{E}{T_sN_t}+(c\hat\tau_{m,n})^{2\beta}$. Therefore, the value of the threshold $\theta$ is linked to the  probability of false alarm $P_{fa}$ as
\begin{eqnarray}
 \nonumber
 P_{fa}=P\left\{\sum_{m=1}^{N_t}\sum_{n=1}^{N_r} \frac{|b_{m,n}|^2}{l_{m,n}}>\theta\;\bigg|\; \mathcal{H}_0\right\}.
\end{eqnarray}
Note that under $\mathcal{H}_0$, $b_{m,n}$ is distributed as ${\cal{CN}}(0,\frac{1}{T_s})$ and therefore, $|b_{m,n}|^2$ is
exponentially distributed with parameter $T_s$. Hence,
\{$|b_{m,n}|^2/l_{m,n}$\} have exponential distributions
with \emph{distinct} parameters $T_s\cdot l_{m,n}^{-1}$. Thus,
$\sum_{m}\sum_n\frac{|b_{m,n}|^2}{l_{m,n}}$ is the sum of $N_tN_r$
independent exponential random variables with different parameters
and its pdf is given by \cite{Balazs:online}
\begin{equation}\label{eq:distribution}
    X\dff \sum_{m}\sum_n\frac{|b_{m,n}|^2}{l_{m,n}}\sim \left[\prod_{m=1}^{N_t}\prod_{n=1}^{N_r}\frac{1}{T_s\cdot l_{m,n}}\right] \sum_{m=1}^{N_t}\sum_{n=1}^{N_r}\frac{\exp\left(-T_s\; l_{m,n}^{-1}\;x\right)}{\prod_{i\neq m}\prod_{j\neq n}T_s(l_{i,j}^{-1}-l_{m,n}^{-1})}.
\end{equation}
If we denote the cumulative distribution function (CDF) of this random variable by $G_1(\cdot)$, then $\theta$ can be determined as $\theta=G_1^{-1}(1-P_{fa})$. By following a similar line of analysis for the case that we deploy the estimate $\hat\btau^{\rm ave}$, the  same detector is found.
\end{proof}

It is noteworthy that our detection scheme has two major differences
with that of [3] provided in [3, Eq. (24)]. First, the detector in [3, Eq. (24)] tests whether a target exists at a {\em known} time-delay while we try to detect a target with {\em unknown} time-delays. Secondly, the model of [3] embeds the effect of the time-delays only as \emph{phase shifts} and ignores the \emph{path-loss} effect. By recalling that the path-losses also depend on the time-delays we have modified the model of [3] to also capture the effects of path-losses.

\section{Joint Detection and Estimation for Point Targets}
\label{sec:point}
In this section we consider the application of MIMO radars for
detecting point targets or exposing single-scatterers. In this case,
the target is modeled by one scatterer located at its gravity center
$X_0$. Note that in (\ref{eq:model_cont}) by setting $P=1$ we get
$h_{m,n}=\zeta\;e^{-j2\pi f_c\tau_{m,n}}$. Unlike in extended targets, the previous section, the distribution of the reflectivity factor $\zeta$ is not
known and thereof it should be treated as an unknown quantity to be
estimated along with the time-delays. Therefore, the null and
alternative hypotheses are given by
\begin{align}\label{eq:single_hypothesisET}
    \left\{\begin{array}{ll}
      \mathcal{H}_0: & r_n[k]=z_{n}[k], \\
      \mathcal{H}_1: &     r_n[k]=\sqrt{\frac{E}{N_t}}\;\zeta\;\sum_{m=1}^{N_t}\frac{c^{-\beta}}{\tau^\beta_{m,n}}\;s_m[k;\tau_{m,n}]\;e^{-j2\pi
    f_c\tau_{m,n}}+z_{n}[k].
    \end{array}\right.
\end{align}
As before, we are interested in solving the optimum test given in
(\ref{eq:GLRT2}) and we start by determining the ML estimates of the
reflectivity $\zeta$ and the time-delay vector $\btau$.

\subsection{Time Delay Estimation}
\label{sec:single_estimation}
Based on the model given in (\ref{eq:single_hypothesisET}) and the
fact that the distribution of $\zeta$ is not known {\em a priori}, we
estimate $\btau$ along with $\zeta$. The pdf of the received vector
$\bR$ for any given $\btau$ and $\zeta$ is $f_1(\bR\med\btau,\zeta)$. By setting $\bx=[\btau,\;\zeta]$ and using the optimal test given in (\ref{eq:GLRT2}) the ML estimate of $\btau$ can be found as shown in the following theorem.
\begin{theorem}\label{th:single_ML}
The ML estimate of $\btau$ for point targets and antenna placement ${\cal X}$ is given by
\begin{equation*}
  \hat\btau=\arg\max_{\btau\in{\cal A}({\cal X})} \frac{\left|\sum_{m=1}^{N_t}\sum_{n=1}^{N_r}\;\frac{e^{j2\pi
  f_c\tau_{m,n}}}{\tau^\beta_{m,n}}\;\sum_{k=1}^{K}r_n[k]\;s^*_m[k;\tau_{m,n}]\right|^2}{\sum_{m=1}^{N_t}\sum_{n=1}^{N_r}\;\frac{1}{\tau^{2\beta}_{m,n}}}.
\end{equation*}
\end{theorem}
\begin{proof}
See Appendix \ref{app:th:point_ML}.
\end{proof}

\subsection{Target Detection}
\label{sec:single_detection}

Based on the ML estimate of $\btau$ provided in Theorem \ref{th:single_ML} the optimal detector is characterized by the following lemma.
\begin{lemma}
\label{lemma:MIMO_point_Det}
The optimal test for point targets and for the given estimate $\hat\btau$ is
\begin{eqnarray*}
  \left|\sum_{m=1}^{N_t}\sum_{n=1}^{N_r}\;\frac{e^{j2\pi  f_c\hat\tau_{m,n}}}{\hat\tau^\beta_{m,n}}\;\sum_{k=1}^{K}r_n[k]\;s^*_m[k;\hat\tau_{m,n}]\right|^2 \underset{\mathcal{H}_0}{\overset{{\cal
  H}_1}{\gtrless}}\theta.
\end{eqnarray*}
\end{lemma}
\begin{proof}
As we are estimating $\btau$ jointly with $\zeta$, by setting $\bx=[\btau,\;\zeta]$ and applying Theorem \ref{th:GLRT} the optimal detector is given by
\begin{eqnarray*}
 \frac{\max_{\bx\in{\cal A}} \pi(\bx)f_{1}(\bR\med\bx)}{f_{0}(\bR)}=\frac{\max_{\btau\in{\cal A}({\cal X}),\zeta} f_{1}(\bR\med\btau,\zeta)}{f_{0}(\bR)}=\frac{f_{1}(\bR\med\hat\btau,\hat\zeta)}{f_{0}(\bR)}
  \underset{\mathcal{H}_0}{\overset{\mathcal{H}_1}\gtrless}\lambda,\quad
\end{eqnarray*}
By following the same line of argument as in Section
\ref{sec:multi_detection} and by using the estimates $\hat\btau$ and $\hat\zeta$ given in Theorem \ref{th:single_ML} and  (\ref{eq:reflecML}), respectively, the optimum test for a point target is given by
\begin{eqnarray}
 \nonumber
  \log\left(\frac{  f_1(\bR\med\boldsymbol{\hat\tau},\hat\zeta\;)}{f_0(\bR)}\right)&=&
  \log\left(\frac{(\pi)^{-KN_r}\exp\left(-\sum_k\|\br[k]\|^2\right) \exp\left(\frac{E}{T_s\;N_t}|\hat{\zeta}\;|^2\;\sum_{m=1}^{N_t}\sum_{n=1}^{N_r}\;\frac{c^{-2\beta}}{\hat\tau^{2\beta}_{m,n}}\right)} {(\pi)^{-KN_r}\exp\left(-\sum_k\|\br[k]\|^2\right)}\right) \\
  &=&\frac{T_s\left|\sum_{m=1}^{N_t}\sum_{n=1}^{N_r}\;\frac{c^{-\beta}}{\hat\tau^\beta_{m,n}}\;e^{j2\pi
  f_c\hat\tau_{m,n}}\;\sum_{k=1}^{K}r_n[k]\;s^*_m[k;\hat\tau_{m,n}]\right|^2}{\sum_{m=1}^{N_t} \sum_{n=1}^{N_r}\;\frac{c^{-2\beta}}{\hat\tau^{2\beta}_{m,n}}}\underset{\mathcal{H}_0}{\overset{{\cal
  H}_1}{\gtrless}}\lambda.
\end{eqnarray}
Moreover, by setting $\theta\dff
\frac{\sum_{m=1}^{N_t}\sum_{n=1}^{N_r}\;{(c\hat\tau_{m,n})^{-2\beta}}}{T_s}\log\lambda$,
the test can be cast as
\begin{eqnarray}
 \nonumber
  \left|\sum_{m=1}^{N_t}\sum_{n=1}^{N_r}\;\frac{e^{j2\pi  f_c\hat\tau_{m,n}}}{\hat\tau^\beta_{m,n}}\;\sum_{k=1}^{K}r_n[k]\;s^*_m[k;\hat\tau_{m,n}]\right|^2 \underset{\mathcal{H}_0}{\overset{{\cal
  H}_1}{\gtrless}}\theta.
\end{eqnarray}

In order to determine the value of the threshold $\theta$, note that
$e^{j2\pi f_c\hat\tau_{m,n}}\sum_{k=1}^{K}r_n[k]\;s^*_m[k;\hat\tau_{m,n}]$ is distributed
as $\mathcal{CN}(0,\frac{1}{T_s})$ under $\mathcal{H}_0$ and
$\sum_{k=1}^{K}r_n[k]\;s^*_m[k;\hat\tau_{m,n}]$ is independent
of $\sum_{k=1}^{K}r_{n'}[k]\;s^*_{m'}[k;\hat\tau_{m',n'}]$ for
$m\neq m'$ or $n\neq n'$. As a result, under $\mathcal{H}_0$ we have
\begin{equation*}
\sum_{m=1}^{N_t}\sum_{n=1}^{N_r}\;\frac{e^{j2\pi
f_c\hat\tau_{m,n}}}{\hat\tau^\beta_{m,n}}\;\sum_{k=1}^{K}r_n[k]\;s^*_m[k;\hat\tau_{m,n}]
\sim
\mathcal{CN}\left(0,\frac{1}{T_s}\sum_{m=1}^{N_t}\sum_{n=1}^{N_r}\frac{1}{\hat\tau_{m,n}^{2\beta}}\right);
\end{equation*}
and consequently
\begin{equation*}
   \left|\sum_{m=1}^{N_t}\sum_{n=1}^{N_r}\;\frac{e^{j2\pi  f_c\hat\tau_{m,n}}}{\hat\tau^\beta_{m,n}}\;\sum_{k=1}^{K}r_n[k]\;s^*_m[k;\hat\tau_{m,n}]\right|^2\sim {\rm
   Exponential}\;\left(\frac{T_s}{\sum_{m=1}^{N_t}\sum_{n=1}^{N_r}\frac{1}{\hat\tau_{m,n}^{2\beta}}}\right).
\end{equation*}
Therefore, for a given value of $P_{fa}$, the threshold level
$\theta$ is found as
\begin{equation*}
    \theta=\frac{1}{T_s}\sum_{m=1}^{N_t}\sum_{n=1}^{N_r}\frac{1}{\hat\tau_{m,n}^{2\beta}}\log\frac{1}{P_{fa}}.
\end{equation*}
\end{proof}

\section{Optimal Joint Estimation/Detection for Phased-array Radar}
\label{sec:PA}

We will compare the performance of MIMO radars against that of
conventional phased-array radar systems. A phased-array system
utilizes an array of closely-located antennas and therefore, the
channel coefficients $h_{m,n}$ are fully correlated, i.e.,
$h_{m,n}=h$. For the case of extended targets $h$ is distributed
as ${\cal CN}(0,1)$ and for point targets its distribution is
unknown \cite{Fishler:SP06}. Moreover, all the emitted waveforms are
equal up to a scaling factor, i.e., $s_m(t)=\hat s_ms(t)$ for
$m=1,\dots,N_t$ such that $\sum_{m=1}^{N_t}|\hat s_m|^2=N_t$ (for
having the total transmitted energy equal to $E$). Therefore, by
using (\ref{eq:model_discrete}) and the narrow-band assumption, the
system model is given by
\begin{equation}\label{eq:model_phased}
    r_n[k]=\sqrt{\frac{E}{N_t}}c^{-\beta}h \;s[k;\tau_{1,1}]\sum_{m=1}^{N_t}\frac{1}{\tau^\beta_{m,n}}\;\hat s_me^{j2\pi f_c(\tau_{1,1}-\tau_{m,n})}+z_{n}[k], \;\;\mbox{for}\;k=1,\dots,K.
\end{equation}
The following lemma is instrumental to deriving the ML estimate of
the time-delay vector.
\begin{lemma}\label{lemma:single}
For any given set of functions $\{g_i(t)\}_{i=1}^N$ where
$g_i(t):\mathbb{R}\rightarrow\mathbb{C}$ and $\alpha,t\in
\mathbb{R}$, we have
\begin{equation*}
 \max_{t_2,\dots,t_N\in \mathbb{R}}\left|\sum_{i=1}^{N}e^{j\alpha
 t_i}g_i(t_1)\right|=\sum_{i=1}^{N}\left|g_i(t_1)\right|.
\end{equation*}
\end{lemma}
\begin{proof}
See Appendix \ref{app:lemma:single}.
\end{proof}
For the purpose of analyzing the diversity gains as given in Section
\ref{sec:diversity}, as well as comparing estimation performance, we
provide the optimal detector and estimator for phased-array radars in the following lemmas. For further analysis we define $S_n(\btau)\dff\sum_{m=1}^{N_t}\frac{1}{(c\tau_{m,n})^\beta}\hat s_me^{j2\pi f_c(\tau_{1,1}-\tau_{m,n})}$ and $\hat S_n(\btau)\dff\sum_{m=1}^{N_t}\frac{1}{(c\tau_{m,n})^\beta}\hat s_me^{j2\pi f_c(\tau_{1,1}-2\tau_{m,n})}$. Similar to ${\cal A}({\cal X})$ which we defined for MIMO radars, we also define ${\cal A}'({\cal X}')$ such that it contains the valid choices of the time-delays for a phased-array radar with closely-located antenna placement ${\cal X}'$.
\begin{lemma}\label{lemma:PA_est}
For a given antenna placement ${\cal X}'$, the ML estimate of the time-delay vector $\btau$ and the optimal detector in phased-array radars are as follows.
\begin{enumerate}
  \item For extended targets through estimating $h$ we have
  \begin{equation*}
  \hat\btau^{\rm MAP}=\arg\max_{\btau\in{\cal A'}({\cal X}')} \frac{\left|\sum_{n=1}^{N_r}S^*_n(\btau)\sum_{k=1}^{K} r_n[k]s^*[k;\tau_{1,1}]\right|^2}
  {\frac{1}{T_s}\sum_{n=1}^{N_r}|S_n(\btau)|^2+1}.
\end{equation*}
  \item For extended targets through averaging over all realizations of $h$ we have
      \begin{equation*}
  \hat\btau^{\rm ave}=\arg\max_{\btau\in{\cal A}'({\cal X}')}\left\{\frac{ \left|\sum_{n=1}^{N_r} S^*_n(\btau) \sum_{k=1}^{K} r_n[k]s^*[k;\tau_{1,1}]\right|^2}{\frac{1}{T_s}\sum_{N=1}^{N_r}|S_n(\btau)|^2+\frac{N_t}{E}}-\log\left(\frac{E}{T_s N_t}\sum_{n=1}^{N_r}|
  S_n(\btau)|^2+1\right)\right\}.
\end{equation*}
\item For point targets through estimating $\zeta$ we have
  \begin{equation*}
  \hat\btau=\arg\max_{\btau\in{\cal A}'({\cal X}')} \frac{\left|\sum_{n=1}^{N_r}\hat S^*_n(\btau)\sum_{k=1}^{K} r_n[k]s^*[k;\tau_{1,1}]\right|^2}
  {\frac{1}{T_s}\sum_{n=1}^{N_r}|\hat S_n(\btau)|^2}.
\end{equation*}
\end{enumerate}
\end{lemma}
\begin{proof}
See Appendix \ref{app:lemma:PAExtended}.
\end{proof}

Based on the ML estimates of $\btau$ provided in Lemma \ref{lemma:PA_est}, the optimal detectors are presented in the
following lemma.
\begin{lemma}
\label{lemma:PA_det}
The optimal test for the given estimate $\btau$ is given as follows.
\begin{enumerate}
  \item For extended targets
  \begin{equation*}
  \bigg|\sum_{n=1}^{N_r}S_n^*(\btau) \sum_{k=1}^{K} r_n[k]s^*[k;\hat\tau_{1,1}]\bigg|^2\underset{\mathcal{H}_0}{\overset{{\cal
  H}_1}{\gtrless}}\hat\theta.
\end{equation*}
  \item For point targets
  \begin{equation*}
  \Big|\sum_{n=1}^{N_r}\hat S_n^*(\btau) \sum_{k=1}^{K} r_n[k]s^*[k;\hat\tau_{1,1}]\Big|^2\underset{\mathcal{H}_0}{\overset{{\cal H}_1}{\gtrless}}\hat\theta.
\end{equation*}
\end{enumerate}
\end{lemma}
\begin{proof} For the case that we are estimating $\btau$ and $h$ jointly, we set $\bx=[\btau,\;h]$ and therefore, the optimal detector is given by
\begin{eqnarray*}
 \frac{\max_{\bx\in{\cal A}} \pi(\bx)f_{1}(\bR\med\bx)}{f_{0}(\bR)}=\frac{\max_{\btau\in{\cal A}'({\cal X}'),\bh} f_{1}(\bR\med\btau,\bh)f_{\bh}(\bh)}{f_{0}(\bR)}=\frac{f_{1}(\bR\med\hat\btau,\hat\bh)f_{\bh}(\hat\bh)}{f_{0}(\bR)}
  \underset{\mathcal{H}_0}{\overset{\mathcal{H}_1}\gtrless}\hat\lambda.
\end{eqnarray*}
Proceeding with the same line of argument as in Section
\ref{sec:multi_detection}, the optimal test is given by
\begin{eqnarray}
 \nonumber
  \frac{f_1(\bR\med{\hat\btau},\hat \bh)f_{\bh}(\hat \bh)}{f_0(\bR)}= \exp\bigg\{\frac{\big|\sum_{n=1}^{N_r}S_n^*(\hat\btau) \sum_{k=1}^{K} r_n[k]s^*[k;\hat\tau_{1,1}]\big|^2}
  {\frac{1}{T_s}\sum_{n=1}^{N_r}|S_n^*(\hat\btau)|^2+\frac{N_t}{E}}\bigg\} \underset{\mathcal{H}_0}{\overset{{\cal H}_1}{\gtrless}}\hat\lambda,
\end{eqnarray}
or equivalently
\begin{equation*}
  \bigg|\sum_{n=1}^{N_r}S_n^*(\hat\btau) \sum_{k=1}^{K} r_n[k]s^*[k;\hat\tau_{1,1}]\bigg|^2\underset{\mathcal{H}_0}{\overset{{\cal
  H}_1}{\gtrless}}\hat\theta,
\end{equation*}
where
\begin{equation*}
   \hat\theta=\log\hat\lambda \cdot \bigg(\frac{1}{T_s}\sum_{n=1}^{N_r}|S_n^*(\hat\btau)|^2+\frac{N_t}{E}\bigg).
\end{equation*}
Under $\mathcal{H}_0$, the term
$S_n^*(\hat\btau)\sum_{k=1}^{K}r_n[k]\;s^*[k;\hat\tau_{1,1}]$ is
distributed as $\mathcal{CN}(0,\frac{| S_n^*(\hat\btau)|^2}{T_s})$
and $\sum_{k=1}^{K}r_n[k]\;s^*[k;\hat\tau_{1,1}]$ is independent of
$\sum_{k=1}^{K}r_{n'}[k]\;s^*[k;\hat\tau_{1,1}]$ for $n\neq n'$.
As a result, under $\mathcal{H}_0$ we have
\begin{equation*}
\sum_{n=1}^{N_r}S_n^*(\hat\btau)\sum_{k=1}^{K}r_n[k]\;s^*[k;\hat\tau_{1,1}]
\sim \mathcal{CN}\left(0,\frac{1}{T_s}\sum_{n=1}^{N_r}|
S_n(\hat\btau)|^2\right);
\end{equation*}
and consequently
\begin{equation*}
   \bigg|\sum_{n=1}^{N_r}S^*_n(\hat\btau)\sum_{k=1}^{K}r_n[k]\;s^*[k;\hat\tau_{1,1}]\bigg|^2\sim {\rm
   Exponential}\;\left(\frac{T_s}{\sum_{n=1}^{N_r}|
S_n(\hat\btau)|^2}\right).
\end{equation*}
Therefore, for a given value of $P_{fa}$, the threshold level
$\theta$ is found by
\begin{equation*}
    \theta=\frac{1}{T_s}\sum_{n=1}^{N_r}|
S_n(\hat\btau)|^2\;\log\frac{1}{P_{fa}}.
\end{equation*}
We can similarly find the same detector when the estimate $\hat\btau^{\rm ave}$ is computed by averaging out the effect of $h$. \end{proof}

\section{Diversity Gain Analysis}
\label{sec:diversity} In order to compare the performance of MIMO
and phased-array radars quantitatively, we analyze the diversity
gain, which determines how fast the mis-detection probability decays
as the transmission $\snr$ increases while there is an upper-bound
constraint on the false-alarm probability~\cite{Haimovich:SPM08}.
More specifically we are interested in finding
\begin{equation*}
    d=-\lim_{\snr\rightarrow\infty}\frac{\log P_{md}(\snr)}{\log\snr},
\end{equation*}
where $P_{md}(\snr)$ denotes the mis-detection probability at the
signal-to-noise ratio $\snr$. The diversity gain for the MIMO radar model provided in \cite{Fishler:SP06} has also been examined in \cite{Fishler:SP06, Haimovich:SPM08}. \cite{Fishler:SP06} also analyzes another notion of diversity, which is the asymptotic value of the detectors's divergence.

Throughout the analysis in the sequel, we say two functions $f_1(x)$ and $f_2(x)$ are \emph{asymptotically equal} when $\lim_{x\rightarrow\infty}\frac{f_1(x)}{f_2(x)}=1$ and denote it by $f_1(x)\circeq f_2(x)$. We also define $\circlt$ and $\circgt$ accordingly. We also say two functions $f_1(x)$ and $f_2(x)$ are \emph{exponentially equal} when $\lim_{x\rightarrow\infty}\frac{\log f_1(x)}{\log f_2(x)}=1$ and denote it by $f_1(x)\doteq f_2(x)$.
\begin{remark}\label{remark:1}
Asymptotic equality is a sufficient condition for exponential equality.
\end{remark}
We will use the following lemma for analyzing the diversity gain of
MIMO and phased-array radars.

\begin{lemma}\label{lemma:diversity}
For any $M$ independent Gaussian random variables
$Y_m\sim\mathcal{N}(\rho\cdot\mu_m,\sigma^2)$, $m=1,\dots,M$, where
$\rho\in\mathbb{R}^+$ and $\mu_m\sim\mathcal{N}(0,\sigma_m^2)$, and
for any given $\gamma\in \mathbb{R}^+$, in the asymptote of large
values of $\rho$ we have
\begin{equation}\label{eq:diversity}
    \bbe_{\bmu}\left[\pr\left(\sum_{m=1}^MY_m^2<\gamma\right)\right]\doteq \rho^{-M},
\end{equation}
where $\bmu\dff[\mu_1,\dots,\mu_M]$.
\end{lemma}
\begin{proof}
See Appendix \ref{app:lemma:diversity}.
\end{proof}
By using the lemma above, in the following theorem we establish the
diversity gain achieved by MIMO and phased-array radars for both
extended and point targets.
$P_{md}(\snr)$.
\begin{theorem}\label{th:MIMO:diversity:extended}
The diversity gain achieved by
\begin{enumerate}
  \item an $N_t\times N_r$ MIMO radar system for extended targets is $N_tN_r$, i.e., $P^{\rm
  E}_{md}(\snr)\doteq\snr^{-N_tN_r}$;
  \item an $N_t\times N_r$ MIMO radar system for point targets is 1, i.e., $P^{\rm
  P}_{md}(\snr)\doteq\snr^{-1}$;
  \item an $N_t\times N_r$ phased array system for extended and point targets is 1, i.e., $P^{\rm PA}_{md}(\snr)\doteq\snr^{-1}$.
\end{enumerate}
\end{theorem}
\begin{proof}
\begin{enumerate}
  \item For extended targets, by recalling the definitions of $b_{m,n}=\sum_{k=1}^{K}r^*_n[k]\;s_m[k;\hat\tau_{m,n}]$ and $l_{m,n}=\frac{E}{T_sN_t}+(c\hat\tau_{m,n})^{2\beta}$, from the optimal test in (\ref{eq:multi_test}) we have \begin{equation}\label{th:MIMO:diversity:extended:proof1}
        P^{\rm E}_{md}(\snr)= \pr\left(\sum_{m=1}^{N_t}\sum_{n=1}^{N_r}\frac{|b_{m,n}|^2}{l_{m,n}} <\theta\;\bigg|\; \mathcal{H}_1\right).
      \end{equation}
      We define $b_{m,n}^r\dff\Re(b_{m,n})$, $b_{m,n}^i\dff\Im(b_{m,n})$. Therefore, the mis-detection probability given in (\ref{th:MIMO:diversity:extended:proof1}) is equivalently given by
      \begin{equation}\label{th:MIMO:diversity:extended:proof2}
        P^{\rm E}_{md}(\snr)= \pr\left(\sum_{m=1}^{N_t}\sum_{n=1}^{N_r}\frac{1}{l_{m,n}}\bigg[(b^r_{m,n})^2+(b^i_{m,n})^2\bigg]<\theta\;\bigg|\; \mathcal{H}_1\right).
      \end{equation}
      By further defining $l_{\min}=\min_{m,n}l_{m,n}$ and $l_{\max}=\max_{m,n}l_{m,n}$ we can provide the following upper and lower bounds on the probability in (\ref{th:MIMO:diversity:extended:proof2}) \begin{align}\label{th:MIMO:diversity:extended:proof3}
        \nonumber &\pr\left(\sum_{m=1}^{N_t}\sum_{n=1}^{N_r}(b^r_{m,n})^2+(b^i_{m,n})^2<\theta l_{\min}\;\bigg|\; \mathcal{H}_1\right)\leq  P^{\rm E}_{md}(\snr)\\
         &\qquad\qquad\qquad\qquad\leq \pr\left(\sum_{m=1}^{N_t}\sum_{n=1}^{N_r}(b^r_{m,n})^2+(b^i_{m,n})^2<\theta l_{\max}\;\bigg|\; \mathcal{H}_1\right).
      \end{align}
      On the other hand, under hypothesis $\mathcal{H}_1$ from (\ref{eq:multi_hypothesisET}) we have
      \begin{equation*}
        b_{m,n} = \sqrt{\frac{E}{N_t}}\;c^{-\beta} \sum_{m'=1}^{N_t}\frac{1}{\tau_{m',n}^\beta}\;h_{m',n} \underset{\frac{1}{T_s}\delta(m-m')}{\underbrace{\sum_{k=1}^Ks_{m'}[k;\tau_{m',n}]\;s_{m}[k;\hat\tau_{m,n}]}}+ \underset{\sim\;\mathcal{CN}(0,\frac{1}{T_s})}{\underbrace{\sum_{k=1}^Kz_{n}[k]s_{m}[k;\hat\tau_{m,n}]}},
      \end{equation*}
      which in turn provides
      \begin{equation*}
         b_{m,n}\sim \mathcal{CN}\left(\sqrt{\frac{E}{N_t}}\;(c\tau_{m,n})^{-\beta}\cdot\frac{h_{m,n}}{T_s},\frac{1}{T_s}\right).
      \end{equation*}
      Furthermore, by defining $h_{m,n}^r\dff\Re(h_{m,n})$ and $h_{m,n}^i\dff\Im(h_{m,n})$ we have $h_{m,n}^r, h_{m,n}^i\sim\mathcal{N}(0,\frac{1}{2})$ and
      \begin{equation*}
        b^r_{m,n}\sim \mathcal{N}\left(\sqrt{\frac{E}{N_t}}\;(c\tau_{m,n})^{-\beta}\cdot\frac{h^r_{m,n}}{T_s},\frac{1}{2T_s}\right)\quad\mbox{and}\quad b^i_{m,n}\sim \mathcal{N}\left(\sqrt{\frac{E}{N_t}}\;(c\tau_{m,n})^{-\beta}\cdot\frac{h^i_{m,n}}{T_s},\frac{1}{2T_s}\right).
      \end{equation*}
      Now, we apply Lemma \ref{lemma:diversity} by setting $M=2N_tN_r$, $\{Y_m\}_m=\{b_{m,n}^r,b_{m,n}^i\}_{m,n}$, $\rho=\sqrt{E}$, $\{\mu_m\}_m=\sqrt{\frac{1}{N_t}}(c\tau_{m,n})^{-\beta}\times\{h^r_{m,n},h^i_{m,n}\}$, $\sigma^2=\frac{1}{2T}$ and $\sigma^2_\mu=\frac{1}{2}$. Thus, we get $\forall\gamma$
      \begin{equation*}
        \pr\left(\sum_{m=1}^{N_t}\sum_{n=1}^{N_r}(b^r_{m,n})^2+(b^i_{m,n})^2<\gamma\;\bigg|\; \mathcal{H}_1\right)\doteq(\sqrt{E})^{-2N_tN_r}
        =(T_s\snr)^{-N_tN_r}\doteq\snr^{-N_tN_r}.
      \end{equation*}

      Hence, for the choices of $\gamma=\theta l_{\min}$ and $\gamma=\theta l_{\max}$ we find that the lower and upper bounds on $P^{\rm E}_{md}(\snr)$ given in (\ref{th:MIMO:diversity:extended:proof3}) are both  exponentially equal to $\snr^{-N_tN_r}$. This results in an identical exponential order for $P^{\rm E}_{md}(\snr)$, i.e.,
      \begin{equation*}
        P^{\rm E}_{md}(\snr)\doteq \snr^{-N_tN_r},
      \end{equation*}
      which establishes the desired result for MIMO radars with extended targets.
      \item We define
      \begin{eqnarray*}
        \tilde b & \dff &  \sum_{m=1}^{N_t}\sum_{n=1}^{N_r}\;\frac{e^{j2\pi  f_c\hat\tau_{m,n}}}{\hat\tau^\beta_{m,n}}\;\sum_{k=1}^{K}r_n[k]\;s^*_m[k;\hat\tau_{m,n}]\\
        & = & \sum_{m=1}^{N_t}\sum_{n=1}^{N_r}\;\frac{e^{j2\pi  f_c\hat\tau_{m,n}}}{\hat\tau^\beta_{m,n}}\sqrt{\frac{E}{N_t}}\;\zeta\;c^{-\beta}\; \sum_{m'=1}^{N_t}\frac{e^{-j2\pi f_c\tau_{m',n}}}{\tau^\beta_{m',n}}\;\underset{=\frac{1}{T_s}\delta(m-m')}{\underbrace{\sum_{k=1}^{K}s_{m'}[k;\tau_{m',n}]s^*_m[k;\hat\tau_{m,n}]}}\\
        & + & \underset{\sim\;\mathcal{CN}\left(0,\;\sum_{m=1}^{N_t}\sum_{n=1}^{N_r}\frac{1}{T_s\hat\tau^{2\beta}_{m,n}}\right)}{\underbrace{\sum_{m=1}^{N_t}\sum_{n=1}^{N_r}\;\frac{e^{j2\pi  f_c\hat\tau_{m,n}}}{\hat\tau^\beta_{m,n}}\sum_{k=1}^Kz_{n}[k]s^*_{m}[k;\hat\tau_{m,n}]}}.
      \end{eqnarray*}
      Therefore, $\tilde b$ is distributed as
      \begin{equation*}
        \tilde b\sim \mathcal{CN}\left(\sqrt{\frac{E}{N_t}}\;c^{-\beta}\sum_{m=1}^{N_t}\sum_{n=1}^{N_r}\;\frac{e^{j2\pi  f_c(\hat\tau_{m,n}-\tau_{m,n})}}{(\hat\tau_{m,n}\tau_{m,n})^\beta}\cdot\frac{\zeta}{T_s},
        \sum_{m=1}^{N_t}\sum_{n=1}^{N_r}\frac{1}{T_s\hat\tau^{2\beta}_{m,n}}\right).
      \end{equation*}
      By following a similar line of argument as for extended targets and defining $\tilde b^r\dff\Re(\tilde b)$, $\tilde b^i\dff\Im(\tilde b)$ and
      \begin{equation*}
        \zeta^r\dff\Re\left(\sum_{m=1}^{N_t}\sum_{n=1}^{N_r}\;\frac{e^{j2\pi  f_c(\hat\tau_{m,n}-\tau_{m,n})}}{(\hat\tau_{m,n}\tau_{m,n})^\beta}\;\zeta\right), \quad \zeta^i\dff\Im\left(\sum_{m=1}^{N_t}\sum_{n=1}^{N_r}\;\frac{e^{j2\pi  f_c(\hat\tau_{m,n}-\tau_{m,n})}}{(\hat\tau_{m,n}\tau_{m,n})^\beta}\;\zeta\right),
      \end{equation*}
      we get
      \begin{equation*}
        \tilde b^r\sim \mathcal{N}\left(\sqrt{\frac{E}{N_t}}\;c^{-\beta}\cdot\frac{\zeta^r}{T_s},\sum_{m=1}^{N_t} \sum_{n=1}^{N_r}\frac{1}{2T_s\hat\tau^{2\beta}_{m,n}}\right),\quad\mbox{and}\quad \tilde b^i\sim \mathcal{N}\left(\sqrt{\frac{E}{N_t}}\;c^{-\beta}\cdot\frac{\zeta^i}{T_s},\sum_{m=1}^{N_t} \sum_{n=1}^{N_r}\frac{1}{2T_s\hat\tau^{2\beta}_{m,n}}\right),
      \end{equation*}
      and
      \begin{equation*}
        P_{md}^{\rm P}(\snr)=\pr\left(|\tilde b|^2 <\theta\right)= \pr\left((\tilde b^r)^2+(\tilde b^i)^2 <\theta\right).
      \end{equation*}
      Now, we apply Lemma \ref{lemma:diversity} by setting $M=2$, $\{Y_m\}_m=\{\tilde b^r,\tilde b^i\}$, $\rho=\sqrt{E}$, $\{\mu_m\}_m=\sqrt{\frac{1}{N_t}}c^{-\beta}\times\{\zeta^r,\zeta^i\}$, $\sigma^2=\frac{1}{2T}$ and $\sigma^2_1=\bbe[(\zeta^r)^2],\;\sigma^2_2=\bbe[(\zeta^i)^2]$. Therefore,
      \begin{equation*}
        P_{md}^{\rm P}(\snr)\doteq(\sqrt{E})^{-2}
        =(T_s\snr)^{-1}\doteq\snr^{-1}.
      \end{equation*}

      \item We provide the proof for extended targets. Similar line of argument and appropriate modifications provide the proof for point targets. Based on the optimal test for detecting extended targets by phased-array radars given in Lemma \ref{lemma:PA_det} we have
      \begin{equation*}
          P_{md}^{\rm PA} =\pr\left(\bigg|{{\sum_{n=1}^{N_r}S^*_n(\hat\btau)\sum_{k=1}^{K}r_n[k]\;s^*[k;\hat\tau_{1,1}]}}\bigg|^2<\theta\;\bigg|\; \mathcal{H}_1\right).
      \end{equation*}
      Under hypothesis $\mathcal{H}_1$ from (\ref{eq:model_phased}) for extended targets we have
      \begin{eqnarray*}
        \hat b\dff \sum_{n=1}^{N_r}S^*_n(\hat\btau)\sum_{k=1}^{K}r_n[k]\;s^*[k;\hat\tau_{1,1}]&=&\sqrt{\frac{E}{N_t}}\; \underset{\dff\;\hat h}{\underbrace{\sum_{n=1}^{N_r}S^*_n(\hat\btau)S_n(\btau)h}}\;\; \underset{\frac{1}{T_s}}{\underbrace{\sum_{k=1}^Ks[k;\tau]s^*[k;\hat\tau]}}\\
        &+& \underset{\sim\;\mathcal{CN}(0,\frac{N_r}{T_s})}{\underbrace{\sum_{n=1}^{N_r}\sum_{k=1}^Kz_{n}[k]s^*[k;\hat\tau]}},
      \end{eqnarray*}
      where $h\sim\;\mathcal{CN}(0,|\sum_{n=1}^{N_r}S^*_n(\hat\btau)S_n(\btau)|^2)$. Therefore,
      \begin{equation*}
        \hat b\sim \mathcal{CN}\left(\sqrt{\frac{E}{N_t}}\cdot\frac{\hat h}{T_s},\frac{N_r}{T_s}\right).
      \end{equation*}
      Similarly as before we define $\hat b^r\dff\Re(\hat b)$, $\hat  b^i\dff\Im(\hat b)$, $\hat h^r\dff\Re(\hat h)$ and $\hat h^i\dff\Im(\hat  h)$, where we have $\hat h^r, \hat h^i\sim\mathcal{CN}(0,\frac{|\sum_{n=1}^{N_r}S^*_n(\hat\btau)S_n(\btau)|^2}{2})$ and
        \begin{equation*}
          \hat b^r\sim \mathcal{CN}\left(\sqrt{\frac{E}{N_t}}\cdot\frac{\hat h^r}{T_s},\frac{N_r}{2T}\right)\quad\mbox{and}\quad \hat b^i\sim \mathcal{CN}\left(\sqrt{\frac{E}{N_t}}\cdot\frac{\hat h^i}{T_s},\frac{N_r}{2T}\right).
        \end{equation*}
        Again we apply Lemma \ref{lemma:diversity} by setting $M=2$, $\{Y_1,Y_2\}_m=\{\hat b^r,\hat b^i\}$, $\rho=\sqrt{E}$,  $\{\mu_1, \mu_2\}_m=\sqrt{\frac{1}{N_t}}(c\tau)^{-\beta}\times\{\hat  h^r,\hat h^i\}$, $\sigma^2=\frac{N_r}{2T}$ and $\sigma^2_m=\frac{N_tN_r}{2}$. Thus, we get $\forall\gamma$ \begin{equation*}
          \pr\left((\hat b^r)^2+(\hat b^i)^2<\gamma\right)\doteq(\sqrt{E})^{-2}=\snr^{-1}.
        \end{equation*}
        As a result we find the following diversity gain achieved by phased-array radar systems for extended targets
        \begin{equation*}
         P^{\rm PA}_{md}(\snr)\doteq \snr^{-1}.
        \end{equation*}
\end{enumerate}
\end{proof}
Although we are using a MIMO radar model different from \cite{Fishler:SP06} and subsequently derive a different detector, the results above conform with those of \cite{Fishler:SP06} which considers detecting the presence of a target at a given location. This result demonstrates that in terms of diversity gain, our proposed MIMO radar model is capable of capturing the same diversity gain achieved by the MIMO radar model of \cite{Fishler:SP06}.

\section{Simulation Results}
\label{sec:simulations}
\subsection{Extended Targets}
\label{sec:simulation_extended}

In this section we provide simulation results on the performance of the proposed joint estimation/detection framework. We consider two antenna configurations with $N_t=N_r=2$ and $N_t=4, N_r=8$. For the MIMO radar we assume that the transmit and receive antennas are located at $X^t_{m}=(m,0,0)$ for $m=1,\dots,4$ and $X^r_n=(0,n,0)$ for $n=1,\dots,8$, respectively. For the phased-array radar we assume that the transmit antennas are all closely-located around $X^t_{m}=(1,0,0)$ for $m=1,\dots,4$, and the receive antennas are closely-located around $X^r_n=(0,1,0)$ for $n=1,\dots,8$. Also we assume that the target to be detected is located at $X_0=(20,15,0)$, where all the distances are in kilometer (km). The path loss coefficient is $\beta=2$ and the carrier frequency is $f_c=5$ MHz. We assume that the target comprises of $P=10$ scatterers and the number of signal samples is $K=40$. Finally, for the MIMO radar the emitted waveforms are $s_{m}(t) = \frac{1}{\sqrt{{T}}}\exp\left(\frac{j2\pi mt}{{T}}\right)\left(U(t)-U(t-{T})\right)$, where $U(t)$ is the unit step function and ${T}$ denotes the duration of the waveform and the sampling rate at the receiver is $T_s=\frac{T}{10}$. For the phased-array radar all the emitted waveforms are equal to $s_{1}(t)$.

\begin{figure}[t!]
  \centering
  \includegraphics[width=4.8 in]{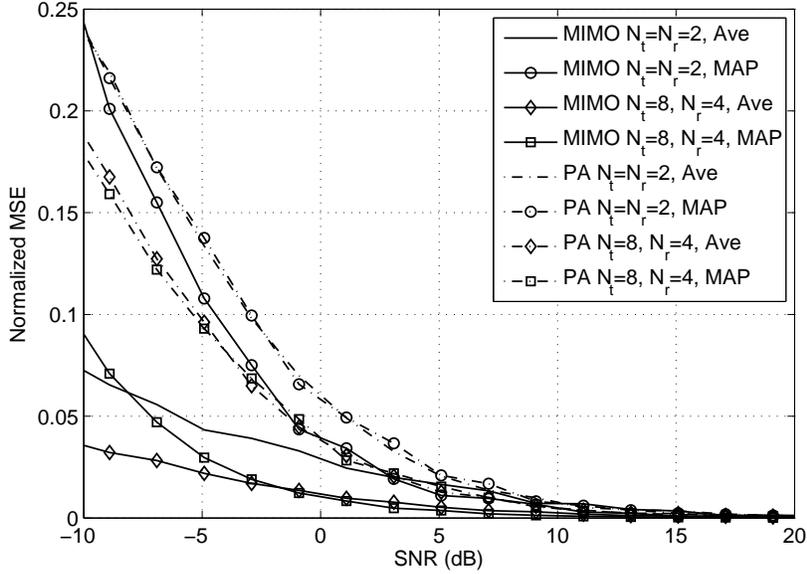}\\
  \caption{Average normalized MSE of time-delay estimates versus $\snr$ for extended targets.}\label{fig:Extended_E}
\end{figure}
\begin{figure}[t!]
  \centering
  \includegraphics[width=4.8 in]{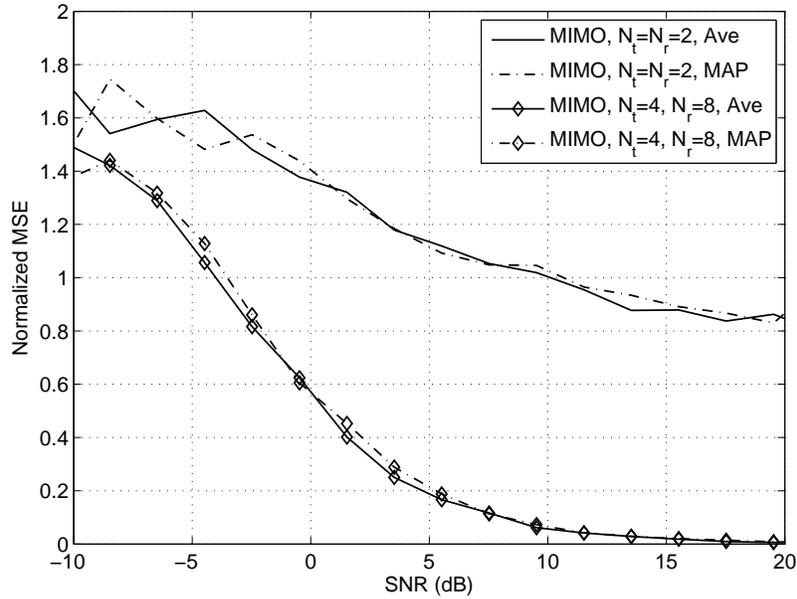}\\
  \caption{Average normalized MSE of location estimates versus $\snr$ for extended targets.}\label{fig:localization}
\end{figure}
\begin{figure}[t!]
  \centering
  \includegraphics[width=4.8 in]{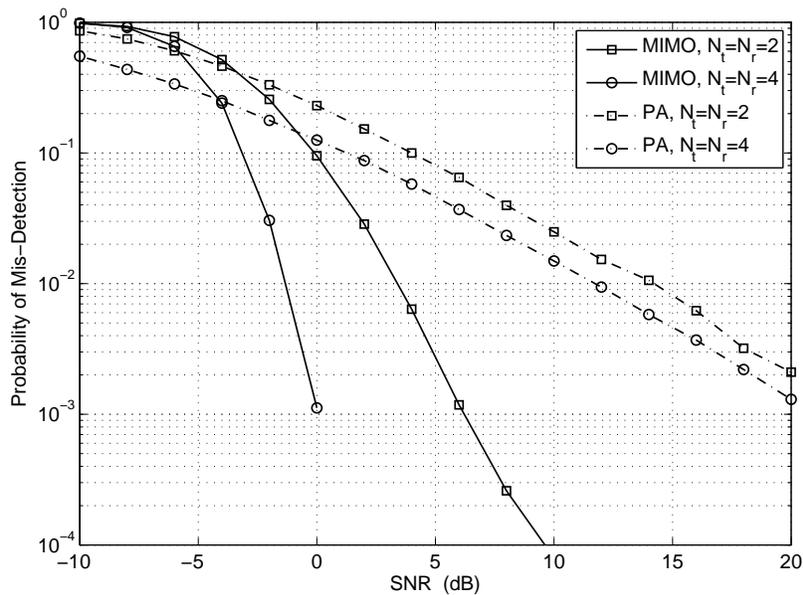}\\
  \caption{Probability of mis-detection versus $\snr$ for a false alarm $P_{fa}=10^{-6}$ for extended targets.}\label{fig:Extended_MD_SNR}
\end{figure}
\begin{figure}[t!]
  \centering
  \includegraphics[width=4.8 in]{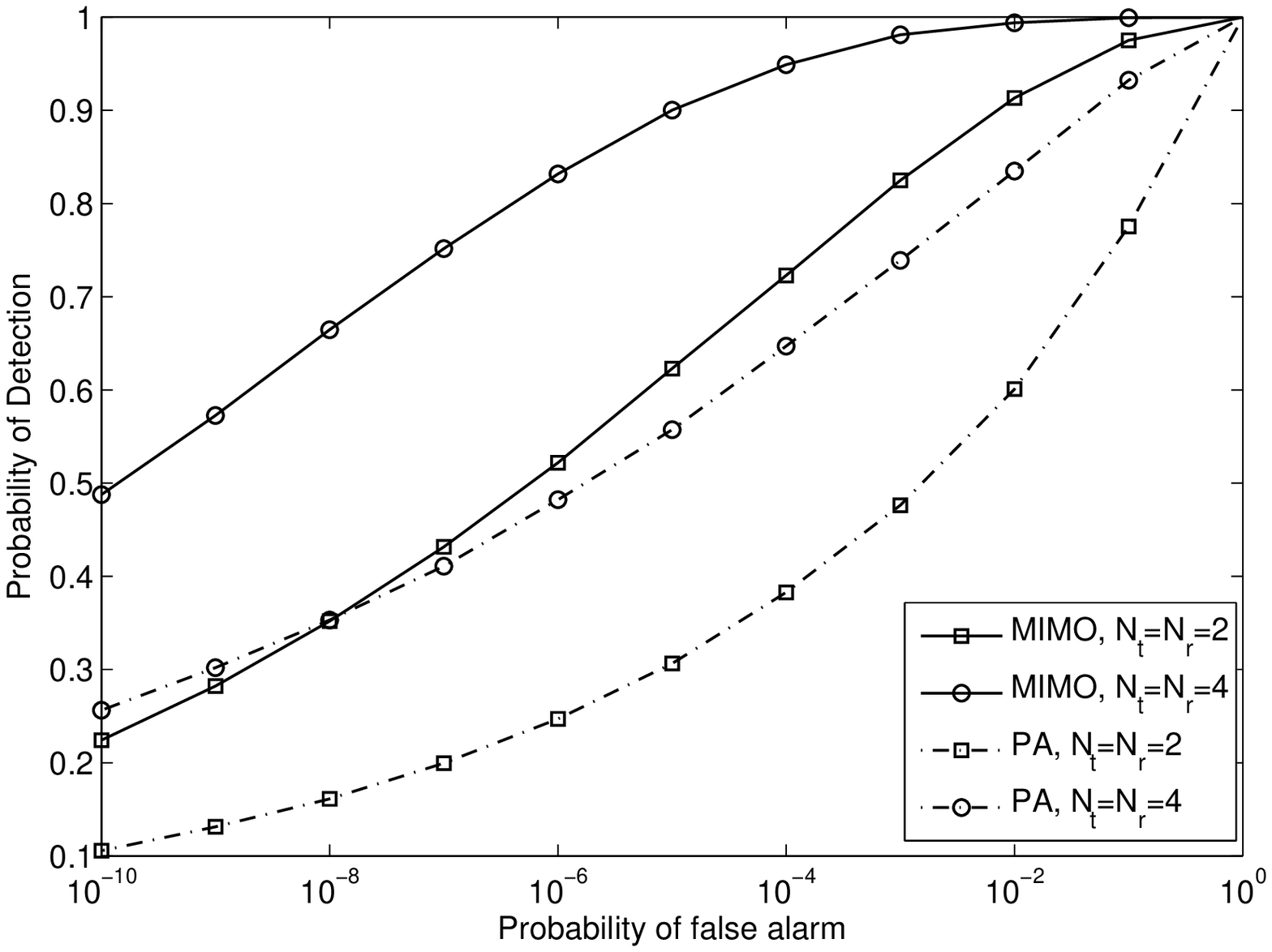}\\
  \caption{Probability of target detection versus probability of false alarm for $\snr = 0$ dB for extended targets.}\label{fig:Extended_D_FA}
\end{figure}
We first consider the performance of parameter estimation. Fig.
\ref{fig:Extended_E} depicts the average normalized mean-square
errors (MSE), i.e., $\frac{1}{N_tN_r}\sum_m\sum_n\big|\frac{\hat\tau_{m,n}-\tau_{m,n}}{\tau_{m,n}}\big|^2$, for phased-array and MIMO radars as a function of received $\snr$. It is observed that the MIMO radar outperforms the  phased-array in all $\snr$ regimes and in particular, by a large margin in  the low $\snr$ regime, which is of more interest in radar applications. Moreover, it is seen that in MIMO radars, the  MAP estimator $\hat\btau^{\rm MAP}$ performs better than the estimator $\hat\btau^{\textrm{ave}}$in the high $\snr$ regime, while the estimator $\hat\btau^{\textrm{ave}}$ outperforms the MAP estimator $\hat\btau^{\textrm{MAP}}$ in the low $\snr$  regime, as expected and discussed in Section~\ref{sec:multi_estimation}.

It should be noted that comparing the accuracies of the time-delay estimators are not as informative as comparing the accuracies of the detectors. This is mainly because the time-delay estimates have different roles in MIMO and phased-array radars. Nevertheless, we perform such comparisons as the detectors accuracies in our formulations depend on the estimators accuracies, which justifies comparing the time-delay estimator in MIMO and phased-array radars.

We next illustrate the localization performance of MIMO radar in
Fig. \ref{fig:localization}. Once the vector $\hat{\btau}$ is  available, it is then possible
to further estimate the location $X_0$ of the gravity center of the
target. The localization problem can be formulated in the form of
the following nonlinear equations
\begin{equation} \label{Eq:Apendix1}
\hat{\btau}=\bphi(X_0)+\bomega,
\end{equation}
where $\bomega$ is the noise term, and
$\bphi(X_0)=\left[\phi_{1,1}(X_0),\dots,\phi_{N_t,N_r}(X_0)\right]$
is the vector of functions defined as
\begin{equation}
  \phi_{m,n}(X_0)\dff\frac{1}{c}\left(\|X_0-X_m^t\|+\|X_0-X_n^r\|\right), m=1,\dots,N_t,\;n=1,\dots,N_r,
\end{equation}
where $X_m^t$ and $X_n^r$ are the positions of the $m^{th}$ and
$n^{th}$ transmit and receive antennas respectively. We can solve
for $X_0$ from (\ref{Eq:Apendix1}) iteratively as follows. Denote
$X^{(i)}_0$ as the solution at the $i^{th}$ iteration. By
linearizing $\bphi(X^{(i+1)}_0)$ with respect to $X^{(i)}_0$ we get
\begin{equation}
  \label{eq:LocIteration}
  \bphi(X^{(i+1)}_0)\approx \bphi(X^{(i)}_0)+\bH_0(X_0^{(i)})\;(X_0^{(i+1)}-X_0^{(i)}),
\end{equation}
where $\bH_0(X_0^{(i)})$ is the Jacobian of the function $\bphi(.)$
evaluated at $X_0^{(i)}$. Then, the least-square estimate of the
position $X_0$ based on (\ref{Eq:Apendix1}) and
(\ref{eq:LocIteration}) is given by
\begin{equation}
X_0^{(i+1)}=\arg\min_{X_0}
\left[\by(X_0^{(i)})-\bH_0(X_0^{(i)})\;X_0\right]^T\left[\by(X_0^{(i)})-\bH_0(X_0^{(i)})\;X_0\right],
\end{equation}
where
$\by(X_0^{(i)})\dff\btau-\left(\bphi(X_0^{(i)})-\bH_0X_0^{(i)}\right)$
and therefore $X^{(i+1)}_0$ has to satisfy the normal equation \begin{equation} \left[\bH_0(X_0^{(i)})^T\bH_0(X_0^{(i)})\right]X^{(i+1)}_0=
\bH_0(X_0^{(i)})^T\by(X_0^{(i)}).
\end{equation}
An initial estimate $X_0^{(0)}$ can be obtained, e.g., using the
method in \cite{Svecova:Radio08}. It is seen that the $4\times 8$ MIMO radar performs considerably better than the $2\times 2$ MIMO radar, which is due to the fact that 32 time-delays provide much more information about the position of the target than 4 time-delays do.

Finally we consider the detection performance. In Fig. \ref{fig:Extended_MD_SNR} the probability of mis-detection versus
$\snr$ is illustrated. The tests are designed such that the
probability of false alarm is $P_{fa}\leq 10^{-6}$. As analyzed in
Section \ref{sec:diversity} and observed in this figure, the slope
of the mis-detection probability of the phased-array radar is 1
decade per 10 dB, whereas that of the MIMO radar is $N_tN_r$ times
steeper. Figure \ref{fig:Extended_D_FA} shows the receiver operating
curve (ROC) for the MIMO and the phased-array systems, for $\snr=0$
dB. It is seen that the MIMO radar significantly outperforms the
phased-array radar over a wide range of false alarm values.
\subsection{Point Targets} \label{sec:simulation_point}
\begin{figure}[t!]
  \centering
  \includegraphics[width=4.8 in]{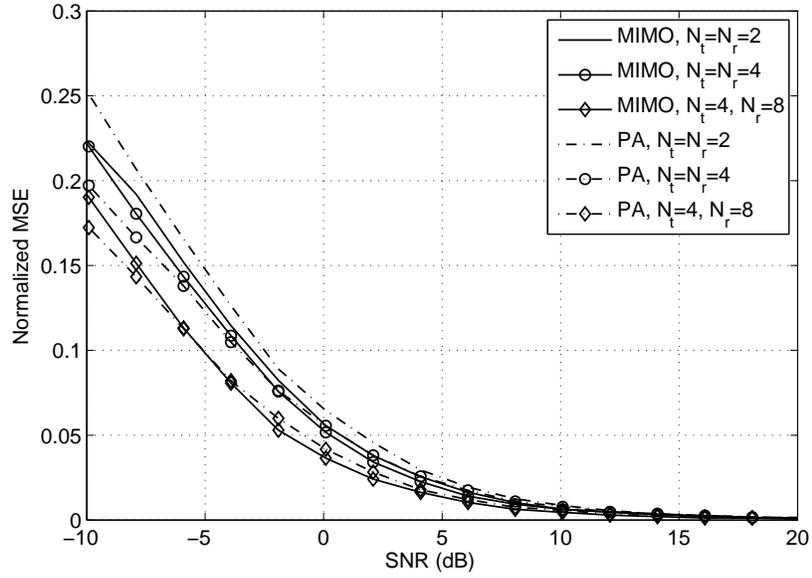}\\
  \caption{Average normalized MSE of time-delay estimates versus $\snr$ for point targets.}\label{fig:Point_E}
\end{figure}
\begin{figure}[t!]
  \centering
  \includegraphics[width=4.8 in]{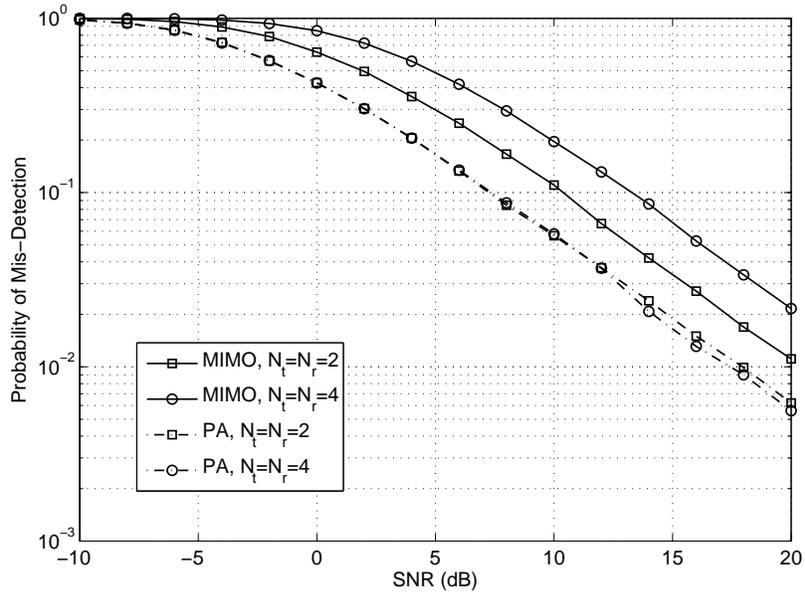}\\
  \caption{Probability of mis-detection versus $\snr$ for point targets and $P_{fa}=10^{-6}$ for point targets.}\label{fig:Point_MD_SNR}
\end{figure}
\begin{figure}[t!]
  \centering
  \includegraphics[width=4.8 in]{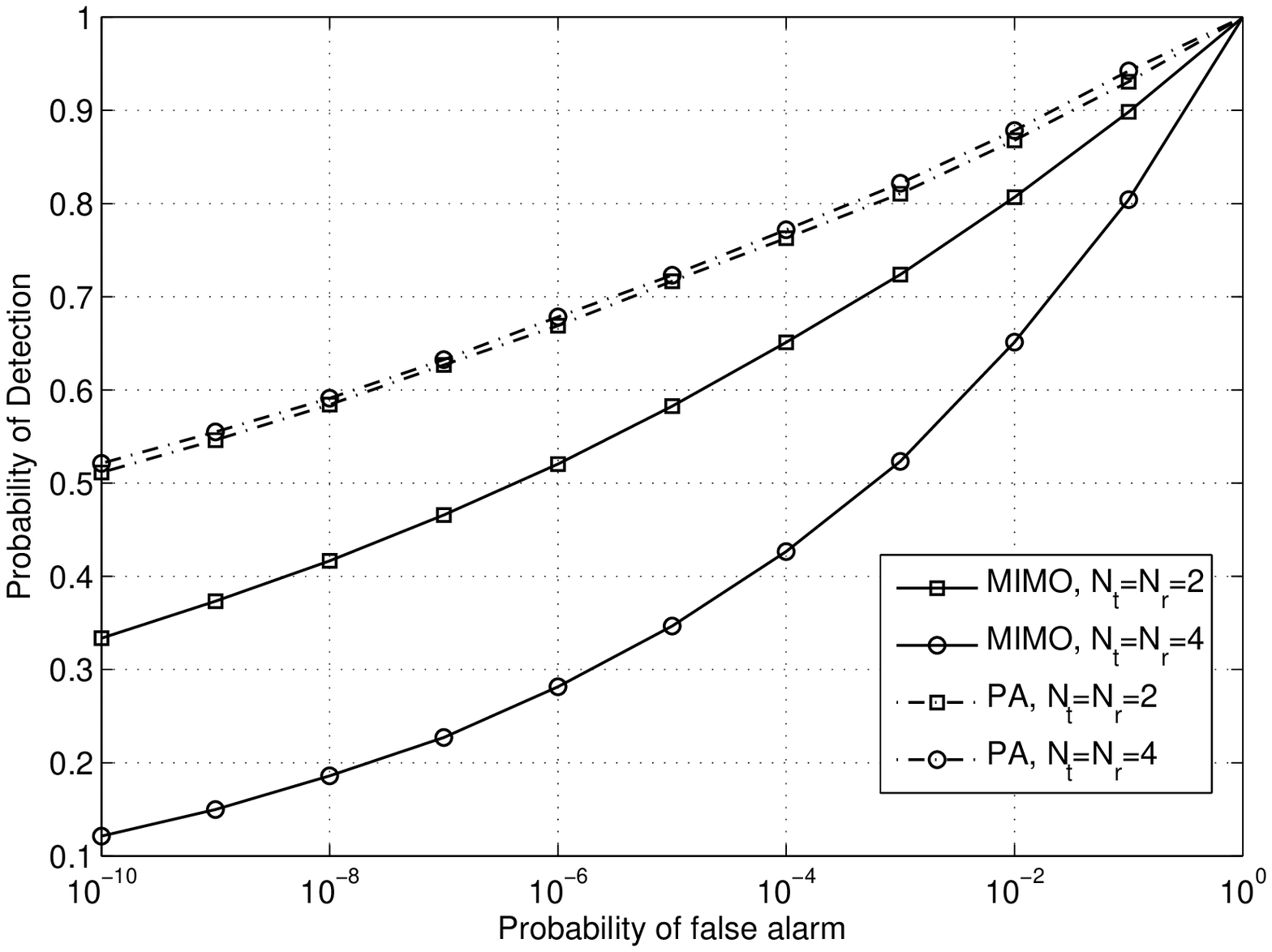}\\
  \caption{Probability of target detection versus probability of false alarm for point targets and $\snr$=0 dB for point targets.}\label{fig:Point_D_FA}
\end{figure}
We use a similar system setup as in Section  \ref{sec:simulation_extended}. Fig. \ref{fig:Point_E} illustrates the normalized average MSE in estimating the time-delays versus $\snr$. Unlike for extended targets, it is observed that for point targets, conventional phased-array and MIMO radars exhibit similar target detection and time-delay estimation accuracies. Therefore, when considering joint target detection and time-delay estimation, deploying MIMO radars are not much advantageous for point targets. This conclusion is nevertheless limited to the specific problem analyzed in this paper and MIMO radars can be potentially advantageous in other scenarios like those discussed in \cite{Haimovich:SPM08} and references therein. This is due to the fact that point targets lack independent scattering section and thereof provide no diversity gain.

In Fig. \ref{fig:Point_MD_SNR} and Fig. \ref{fig:Point_D_FA} the
probability of mis-detection versus $\snr$ and the ROC are plotted, respectively. For Fig. \ref{fig:Point_MD_SNR} the tests are designed such that the probability of false alarm is $P_{fa}\leq 10^{-6}$ and for Fig. \ref{fig:Point_D_FA} we have set $\snr$=0 dB. It is seen from
Fig. \ref{fig:Point_MD_SNR} that both the phase-array and the MIMO
radars exhibit a diversity gain of 1, which verifies Theorem
\ref{th:MIMO:diversity:extended}.

\section{Conclusions}
\label{sec:conclusions}

In this paper we have offered a framework for jointly detecting the presence of a target and estimating some unknown target parameters. We have provided a composite hypothesis test that satisfies some optimality criteria for both target detection and parameter estimation. By using this framework we have proposed a new MIMO radar model for detecting the presence of a target in an unknown location and formulated a composite hypothesis test for solving this problem. In this problem, the unknown parameter of interest to be jointly estimated along with target detection, is the vector of time-delays that a transmitted waveform  experiences from being emitted by the transmit antennas until being received by the receive antennas. For the proposed detection scheme, we also have analyzed the decay rate of the probability of mis-detection with increasing $\snr$ while the false-alarm probability is kept below a certain level (diversity gain). Our simulation results demonstrate that for the specific problem of joint target detection and time-delay estimation, deploying MIMO radars is advantageous for extended targets. On the other hand, for point targets, the accuracies of target detection and time-delay estimation by conventional phased-array are comparable with those yielded by MIMO radars.

\appendix

\section{Proof of Theorem \ref{th:multi_ML}}
\label{app:th:multi_ML}
The pdf of the vector of received signals $\bR$ for any given $\btau$ and $\bh$ is given by
\begin{align}
  \label{eq:multi_pdf_H1_1}
  f_1(\bR\med {\boldsymbol{\tau}},\bh)&=(\pi)^{-KN_r}
  \exp\Bigg(-\underset{\dff D_1(\btau,\bh)}{\underbrace{\sum_{n=1}^{N_r}\sum_{k=1}^{K}\left|r_n[k]- \sqrt{\frac{E}{N_t}}\;c^{-\beta}\;\sum_{m=1}^{N_t}\frac{1}{\tau^\beta_{m,n}}\; h_{m,n}\;s_m[k;\tau_{m,n}]\right|^2}}\Bigg).
\end{align}
By taking into account the orthogonality assumption
$\sum_{k=1}^{K}s_m[k;\tau_{m,n}]\;s^*_{m'}[k;\tau_{m',n}]=\frac{\delta(m-m')}{T_s}$,
we can further simplify the term $D_1(\btau,\bh)$ defined in (\ref{eq:multi_pdf_H1_1}) as follows.
\begin{align}
  \nonumber D_1(\btau,\bh)
  &=\sum_{n=1}^{N_r}\sum_{k=1}^{K}\left|r_n[k]-\sqrt{\frac{E}{N_t}}\;c^{-\beta}\; \sum_{m=1}^{N_t}\frac{1}{\tau^\beta_{m,n}}\;h_{m,n}\;s_m[k;\tau_{m,n}]\right|^2\\
  \nonumber &=\sum_{k=1}^{K}\|\br[k]\|^2+\frac{E}{N_t}\;c^{-2\beta}\;\sum_{n=1}^{N_r}\ \sum_{m=1}^{N_t}\sum_{m'=1}^{N_t}\frac{1}{\tau^\beta_{m,n}\tau^\beta_{m',n}}\;h_{m,n}\;h^*_{m',n}\; \underset{\frac{1}{T_s}\;\delta(m-m')}{\underbrace{\sum_{k=1}^{K}s_m[k;\tau_{m,n}]\;s^*_{m'}[k;\tau_{m',n}]}}\\
  \nonumber
  &-\sqrt{\frac{E}{N_t}}\;c^{-\beta}\sum_{n=1}^{N_r}\sum_{m=1}^{N_t}h_{m,n}\;\frac{1}{\tau^\beta_{m,n}}\; \sum_{k=1}^{K}r^*_n[k]\;s_m[k;\tau_{m,n}]\\
  \nonumber  &-\sqrt{\frac{E}{N_t}}\;c^{-\beta}\sum_{n=1}^{N_r}\sum_{m=1}^{N_t}h^*_{m,n}\; \frac{1}{\tau^\beta_{m,n}}\;\sum_{k=1}^{K}r_n[k]\;s^*_m[k;\tau_{m,n}]\\
  \nonumber
  &=\sum_{k=1}^{K}\|\br[k]\|^2+\frac{E}{T_sN_t}\;c^{-2\beta}\sum_{n=1}^{N_r} \sum_{m=1}^{N_t}\frac{1}{\tau^{2\beta_{m,n}}}\;|h_{m,n}|^2\\
  \nonumber
  &-\sum_{n=1}^{N_r}\sum_{m=1}^{N_t}h_{m,n}\;\underset{\dff\; a^*_{m,n}(\tau_{m,n})}{\underbrace{\sqrt{\frac{E}{N_t}}\;\frac{c^{-\beta}}{\tau^\beta_{m,n}} \sum_{k=1}^{K}r^*_n[k]\;s_m[k;\tau_{m,n}]}}\\
  \label{eq:multi_pdf_H1_2}
  &-\sum_{n=1}^{N_r}\sum_{m=1}^{N_t}h^*_{m,n}\;\underset{a_{m,n}(\tau_{m,n})} {\underbrace{\sqrt{\frac{E}{N_t}}\;\frac{c^{-\beta}}{\tau^\beta_{m,n}}
  \sum_{k=1}^{K}r_n[k]\;s^*_m[k;\tau_{m,n}]}}.
\end{align}
For any given time-delay vector $\btau$, define
\begin{equation}\label{eq:multi_pdf_H1_3}
\bA_{\btau}\dff
\frac{E}{T_sN_t}\;c^{-2\beta}\textrm{diag}(\tau^{-2\beta}_{1,1},
\tau^{-2\beta}_{1,2},...,\tau^{-2\beta}_{N_t,N_r})\quad {\rm
and}\quad
\ba_{\btau}\dff[a_{1,1}(\tau_{1,1}),a_{1,2}(\tau_{1,2}),\dots,a_{N_t,N_r}(\tau_{N_t,N_r})]^T.
\end{equation} Then, $D_1(\btau,\bh)$ can be represented in the matrix form as
\begin{eqnarray}
  \label{eq:D_1} D_1(\btau,\bh) &=&\sum_{k=1}^K\|\br[k]\|^2 +\bh^H\bA_{\btau}\;\bh-\bh^H\ba_{\btau}-\ba_{\btau}^H\bh.
\end{eqnarray}
By noting that the pdf of $\bh$ is given by
$f_\bh(\bh)=(\pi)^{-N_tN_r}\exp({-\|\bh\|^2})$, we find the estimates of $\btau$ as follows.

\begin{enumerate}
  \item For finding $\hat\btau^{\rm MAP}$ by denoting the MAP estimate of $\bh$ by $\hat\bh$, setting $\bx=[\hat\btau^{\rm MAP},\;\hat\bh]$, and considering the independence of $\btau$ and $\bh$, from (\ref{eq:GLRT2}) we get
      \begin{equation*}
        [\hat\btau^{\rm MAP};\;\hat\bh]=\arg\max_{\btau\in{\cal A}({\cal X}),\bh} f_1(\bR\med \btau,\bh)f_\bh(\bh)= \arg\max_{\btau\in{\cal A}({\cal X})}\arg\max_{\bh}f_1(\bR\med \btau,\bh)f_\bh(\bh),
      \end{equation*}
      or equivalently
      \begin{equation}\label{eq:h_MAP}
        \hat\bh=\arg\max_{\bh}f_1(\bR\med \btau,\bh)f_\bh(\bh),\quad\mbox{and}\quad \hat\btau^{\rm MAP}=\arg\max_{\btau\in{\cal A}({\cal X})} f_1(\bR\med \btau,\hat\bh)f_\bh(\hat\bh).
      \end{equation}
      Using (\ref{eq:multi_pdf_H1_1}), (\ref{eq:D_1}), and (\ref{eq:h_MAP}) and noting the definition of $\ba_{\btau}$  given in (\ref{eq:multi_pdf_H1_3}) provides that
      \begin{eqnarray}
        \nonumber\hat\bh &=&  \arg\max_{\bh}f_1(\bR\med \btau,\bh)f_\bh(\bh)\\
        \nonumber &=& \arg\max_{\bh}(\pi)^{-N_tN_r+KN_r} \exp\left(-\left(\sum_k\|\br[k]\|^2+\bh^H\bA_{\btau}\bh- \bh^H\ba_{\btau}-\ba_{\btau}^H \bh+\|\bh\|^2\right)\right)\\
        \label{eq:B} &=& \arg\min_{\bh}\bigg\{\bh^H\underset{\dff \bB_{\btau}}{\underbrace{\left(\bA_{\btau}+\bI_{N_tN_r}\right)}}\bh- \bh^H \ba_{\btau}-\ba_{\btau}^H \bh\bigg\}\\
         \nonumber&=&\bB_{\btau}^{-1}\ba_{\btau}.
      \end{eqnarray}
      Based on the above estimate of $\btau$, $D_1(\btau,\hat\bh)$ given in (\ref{eq:D_1}) becomes \begin{eqnarray}
        \label{eq:multi_pdf_H1_4} D_1(\btau,\hat\bh) &=&\sum_{k=1}^K\|\br[k]\|^2 -\ba_{\btau}^H\bB^{-1}\ba_{\btau}-\|\hat\bh\|^2.
      \end{eqnarray}
      Hence, the ML estimate of $\btau$ can be found by solving \begin{align}
          \nonumber\hat\btau^{\rm MAP}=\arg\max_{\btau\in{\cal A}({\cal X})}\log f_1(\bR\med \boldsymbol{\tau},\hat\bh)f_{\bh}(\hat\bh)&= \arg\max_{\btau\in{\cal A}({\cal X})}\left\{-D_1(\btau,\hat\bh)-\|\hat\bh\|^2\right\}\\ &=\arg\max_{\btau\in{\cal A}({\cal X})}\left\{\ba_{\btau}^H\bB^{-1}_{\btau}\ba_{\btau}\right\}\\ \nonumber&=\arg\max_{\btau\in{\cal A}({\cal X})}\left\{\ba_{\btau}^H(\bA_{\btau}+\bI)^{-1}\ba_{\btau}\right\}\\ \label{eq:multi_pdf_H1_5}  &=\arg\max_{\btau\in{\cal A}({\cal X})}\left\{\sum_{m=1}^{N_t}\sum_{n=1}^{N_r} \frac{|a_{m,n}(\tau_{m,n})|^2}{\frac{E}{T_sN_t}(c\tau_{m,n})^{-2\beta}+1}\right\},
      \end{align}
      which concludes the proof.
  \item For this case which we average out the effect of $\bh$ by setting $\bx=\btau$ from (\ref{eq:GLRT2}) we get
      \begin{eqnarray}
       \nonumber
       \hat\btau^{\rm ave}&=&\arg\max_{\btau\in{\cal A}({\cal X})}f_1(\bR\med {\btau})= \arg\max_{\btau\in{\cal A}({\cal X})} \int_\bh f_1(\bR \med {\bold  \tau},\bh)f_\bh(\bh)\;d\bh\\
       \label{eq:multi_pdf_H1_5}
       &=&\arg\max_{\btau\in{\cal A}({\cal X})}\int_\bh \exp\Big(-D_1(\btau,\bh)\Big)f_\bh(\bh)\;d\bh\\
       \nonumber &=&\arg\max_{\btau\in{\cal A}({\cal X})}\int_\bh \exp\bigg(-\Big(\bh^H\underset{\dff \bB_{\btau}}{\underbrace{\left(\bA_{\btau}+\bI_{N_tN_r}\right)}}\bh- \bh^H \ba_{\btau}-\ba_{\btau}^H \bh\Big)\bigg)\;d\bh\\  \label{eq:multi_pdf_H1_6} &=& \arg\max_{\btau\in{\cal A}({\cal X})}\int_\bh \exp\left({-\left\{\left(\bh-\bB_{\btau}^{-1}\ba_{\btau}\right)^H\bB_{\btau}\; \left(\bh-\bB_{\btau}^{-1}\ba_{\btau}\right)-\ba_{\btau}^H\bB_{\btau}^{-1}\ba_{\btau}\right\}}\right)\;d\bh\\ \nonumber &=& \arg\max_{\btau\in{\cal A}({\cal X})}\exp\left(\ba_{\btau}^H\bB_{\btau}^{-1}\ba_{\btau}\right) \int_\bh  \exp\left({-\left\{\left(\bh-\bB_{\btau}^{-1}\ba_{\btau}\right)^H\bB_{\btau}\; \left(\bh-\bB_{\btau}^{-1}\ba_{\btau}\right)\right\}}\right)\;d\bh\\ \label{eq:multi_pdf_H1_7} &= &\arg\max_{\btau\in{\cal A}} \exp\left(\ba_{\btau}^H\bB_{\btau}^{-1}\ba_{\btau}\right)\;|\bB_{\btau}|^{-1}\\ \nonumber &=& \arg\max_{\btau\in{\cal A}({\cal X})}\left\{\ba_{\btau}^H\bB_{\btau}^{-1}\ba_{\btau}-\log|\bB_{\btau}|\right\}\\ \nonumber &=&\arg\max_{\btau\in{\cal A}({\cal X})}\left\{\ba_{\btau}^H(\bA_{\btau}+\bI)^{-1}\ba_{\btau}-\log|\bA_{\btau}+\bI|\right\}\\ \nonumber &=&\arg\max_{\btau\in{\cal A}({\cal X})}\sum_{m=1}^{N_t}\sum_{n=1}^{N_r} \left\{\frac{|a_{m,n}(\tau_{m,n})|^2}{\frac{E}{T_sN_t}(c\tau_{m,n})^{-2\beta}+1}- \log\left(\frac{E}{T_sN_t}(c\tau_{m,n})^{-2\beta}+1\right)\right\}.
  \end{eqnarray}
  Noting the definition of $a_{m,n}(\tau_{m,n})$ given in (\ref{eq:multi_pdf_H1_2}) concludes the proof.
\end{enumerate}

\section{Proof of Theorem \ref{th:single_ML}}
\label{app:th:point_ML}
In order to find the ML estimate of $\btau$ and $\zeta$ form (\ref{eq:GLRT2}) we have
\begin{equation*}
    [\btau,\;\zeta]=\arg\max_{\btau\in{\cal A}({\cal X}),\zeta}f_1(\bR\med\btau,\zeta)=\arg\max_{\btau\in{\cal A}({\cal X})}\arg\max_{\zeta}f_1(\bR\med\btau,\zeta),
\end{equation*}
or equivalently
\begin{equation*}
    \hat\zeta=\arg\max_{\zeta}f_1(\bR\med\btau,\zeta)\quad\mbox{and}\quad \hat\btau=\arg\max_{\btau\in{\cal A}({\cal X})}f_1(\bR\med\btau,\hat\tau).
\end{equation*}
On the other hand, from (\ref{eq:single_hypothesisET}) the likelihood $f_1(\bR\med\btau,\hat\tau)$ is given by
\begin{align}
  \label{eq:single_pdf_H1_1}
  f_1(\bR \med \boldsymbol{\tau},\zeta) =(\pi)^{-KN_r} \exp\Bigg(-\underset{\dff D_2(\btau,\zeta)}{\underbrace{\sum_{n=1}^{N_r}\sum_{k=1}^{K}
  \bigg|r_n[k]-\sqrt{\frac{E}{N_t}}\;\zeta \; c^{-\beta}\sum_{m=1}^{N_t}\frac{e^{-j2\pi f_c\tau_{m,n}}}{\tau^\beta_{m,n}}\;s_m[k;\tau_{m,n}]\bigg|^2}}\Bigg).
\end{align}
By taking into account the orthonormality assumption on $s_m[k;\tau_{m,n}]$ we get
\begin{eqnarray}
\nonumber D_2(\btau,\zeta) &=&
 \sum_{n=1}^{N_r}\sum_{k=1}^{K}\bigg|r_n[k]-\sqrt{\frac{E}{N_t}}\;\zeta \; c^{-\beta}\sum_{m=1}^{N_t}\frac{e^{-j2\pi f_c\tau_{m,n}}}{\tau^\beta_{m,n}}\;s_m[k;\tau_{m,n}]\bigg|^2\\ \nonumber
 &=&\sum_{k=1}^{K}\|\br[k]\|^2+\frac{E}{T_sN_t}\left|\zeta\right|^2c^{-2\beta}\; \sum_{n=1}^{N_r}\sum_{m=1}^{N_t}\frac{1}{(c\tau_{m,n})^{2\beta}}\\ \nonumber &-&\sum_{n=1}^{N_r}\sum_{m=1}^{N_t}e^{-j2\pi f_c\tau_{m,n}}\;{{\sqrt{\frac{E}{N_t}}\;\zeta\; \frac{c^{-\beta}}{\tau^\beta_{m,n}}\;\sum_{k=1}^{K}r^*_n[k]\;s_m[k;\tau_{m,n}]}}\\  &-&\sum_{n=1}^{N_r}\sum_{m=1}^{N_t}e^{j2\pi f_c\tau_{m,n}}\;{{\sqrt{\frac{E}{N_t}}\;\zeta^*\; \frac{c^{-\beta}}{\tau^\beta_{m,n}}\;\sum_{k=1}^{K}r_n[k]\;s^*_m[k;\tau_{m,n}]}}. \label{eq:exponent}
\end{eqnarray}
$D_2(\btau,\zeta)$ is quadratic in $\zeta$ and its minimum which is attained at the ML estimate of $\zeta$ is given by
\begin{align}
\nonumber \hat\zeta=\arg\max_{\zeta}\log f_1(\bR\med {\boldsymbol{\tau}},\zeta) &=\arg\min_{\zeta}\; D_2(\btau,\zeta)\\
\label{eq:reflecML}&=\frac{\sum_{m=1}^{N_t}\sum_{n=1}^{N_r}\;\frac{1}{(c\tau_{m,n})^\beta}\;e^{j2\pi
f_c\tau_{m,n}}\;\sum_{k=1}^{K}r_n[k]\;s^*_m[k;\tau_{m,n}]}{\frac{1}{T_s}\sqrt{\frac{E}{N_T}} \sum_{m=1}^{N_t}\sum_{n=1}^{N_r}\;\frac{1}{(c\tau_{m,n})^{2\beta}}}.
\end{align}
By substituting (\ref{eq:reflecML}) back in (\ref{eq:exponent}),
and after some manipulation, $D_2(\btau,\hat\zeta)$ becomes
\begin{eqnarray}
\nonumber D_2(\btau,\hat\zeta) &=&
\sum_{k=1}^{K}\|\br[k]\|^2-
\frac{E}{T_sN_t}|\hat{\zeta}\;|^2\;\sum_{m=1}^{N_t}\sum_{n=1}^{N_r}\;\frac{1}{(c\tau_{m,n})^{2\beta}}.
\end{eqnarray}
Therefore,
\begin{align}
\nonumber \hat\btau=\arg\max_{\btau\in{\cal A}({\cal X})}\log f_1(\bR\med {\boldsymbol{\tau}},\hat\zeta) &=\arg\min_{\zeta}\; D_2(\btau,\hat\zeta)=\arg\max_{\btau\in{\cal A}({\cal X})} \Big\{|\hat{\zeta}\;|^2\;\sum_{m=1}^{N_t}\sum_{n=1}^{N_r}\;\frac{1}{(c\tau_{m,n})^{2\beta}}\Big\}\\  \label{eq:th_single_ML_1}&=\arg\max_{\btau\in{\cal A}({\cal X})} \frac{\left|\sum_{m=1}^{N_t}\sum_{n=1}^{N_r}\;\frac{e^{j2\pi
f_c\tau_{m,n}}}{\tau^\beta_{m,n}}\;\sum_{k=1}^{K}r_n[k]\;s^*_m[k;\tau_{m,n}]\right|^2} {\sum_{m=1}^{N_t}\sum_{n=1}^{N_r}\;\frac{1}{\tau^{2\beta}_{m,n}}}.
\end{align}

\section{Proof of Lemma \ref{lemma:single}}
\label{app:lemma:single} We provide the proof by induction.
\begin{enumerate}
  \item $N=2$. We define
  \begin{eqnarray}
   \nonumber q(t_1,t_2)\dff \left|e^{j\alpha t_1}g_1(t_1)+e^{j\alpha t_2}g_2(t_1)\right|.
  \end{eqnarray}
  Then
  \begin{eqnarray}
   \nonumber q^2(t_1,t_2)= \left|g_1(t_1)\right|^2+\left|g_2(t_1)\right|^2+e^{j\alpha(t_1-t_2)}g_1(t_1)g_2^*(t_1)+e^{j\alpha(t_2-t_1)}g_2(t_1)g_1^*(t_1),
  \end{eqnarray}
  and therefore,
  \begin{eqnarray}
   \nonumber \frac{\partial(q^2(t_1,t_2))}{\partial t_2}=j\alpha\left[e^{j\alpha(t_2-t_1)}g_2(t_1)g_1^*(t_1)-e^{j\alpha(t_1-t_2)}g_1(t_1)f^*_2(t_1)\right].
  \end{eqnarray}
  For finding $\max_{t_2}q(t_1,t_2)$ we set $\frac{\partial(q^2(t_1,t_2))}{\partial t_2}=0$ and obtain
  \begin{equation*}
    e^{j\alpha\hat{t}_2}=e^{j\alpha{t}_1}\sqrt{\frac{g_1(t_1)g^*_2(t_1)}{g^*_1(t_1)g_2(t_1)}}.
  \end{equation*}
  We can check that this is a maximum by noting that
  \begin{eqnarray*}
   \left.\frac{\partial^2(q^2(t_1,t_2))}{\partial t_2^2}\right|_{t_2=\hat{t}_2}&=&(j\alpha)^2\left[ e^{j\alpha(\hat{t}_2-t_1)}g^*_1(t_1)g_2(t_1) + e^{j\alpha(t_1-\hat{t}_2)}g_1(t_1)g^*_2(t_1)
   \right]\\
   &=&-2(\alpha)^2\left|g_1(t_1)g_2(t_1)\right|< 0,
 \end{eqnarray*}
 Hence,
 \begin{equation*}
   \max_{t_2} q(t_1,t_2) =q(t_1,\hat{t}_2)=\sqrt{|g_1(t_1)|^2+|g_2(t_1)|^2+2|g_1(t_1)g_2(t_1)|}=\sum_{i=1}^2|g_i(t_1)|.
 \end{equation*}
\item Inductive assumption.
 \begin{equation*}
   \max_{\{t_2,\dots,t_N\}}\left|\sum_{i=1}^{N}e^{j\alpha t_i}g_i(t_1)\right|=\sum_{i=1}^{N}\left|g_i(t_1)\right|.
 \end{equation*}
\item Claim
 \begin{equation*}
   \nonumber
   \max_{\{t_2,\dots,t_{N+1}\}}\left|\sum_{i=1}^{N+1}e^{j\alpha t_i}g_i(t_1)\right|=\sum_{i=1}^{N+1}\left|g_i(t_1)\right|.
 \end{equation*}
 Let ${\bt}\dff\left[t_1\;\hdots t_{N+1}\right]$ and $q({\bt})\dff
 \left|\sum_{i=1}^{N+1}e^{j\alpha t_i}g_i(t_1)\right|$, then
 \begin{eqnarray}
   \nonumber
   \frac{\partial(q^2(\bt))}{\partial t_{N+1}}&=&\frac{\partial}{\partial
   t_{N+1}}\left[\sum_{i=1}^{N}e^{j\alpha(t_{N+1}-t_i)}g_i^*(t_1)g_{N+1}(t_1)+\sum_{i=1}^{N}e^{j\alpha(t_i-t_{N+1})}g_i(t_1)g^*_{N+1}(t_1)\right]\\
   \nonumber
   &=&j\alpha\;\left[\sum_{i=1}^{N}e^{j\alpha(t_{n+1}-t_i)}g_i^*(t_1)g_{N+1}(t_1)+\sum_{i=1}^{N}e^{j\alpha(t_i-t_{N+1})}g_i(t_1)g^*_{N+1}(t_1)\right].
 \end{eqnarray}
 If we set $\frac{\partial(q^2(\bt))}{\partial t_{n+1}}=0$, then
 \begin{eqnarray}
   \nonumber
   e^{2j\alpha\hat{t}_{N+1}}=\frac{\sum_{i=1}^{N}e^{j\alpha t_i}g_i(t_1)}{\sum_{i=1}^{N}e^{-j\alpha t_i}g^*_i(t_1)}\frac{g^*_{N+1}(t_1)}{g_{N+1}(t_1)}.
 \end{eqnarray}
 Also $\hat t_{N+1}$ is a maximum since
 \begin{eqnarray}
  \nonumber
  \left.\frac{\partial^2(q^2(\bt))}{\partial t_{N+1}^2}\right|_{t_{N+1}=\hat{t}_{N+1}}&=&(j\alpha)^2\left[ \sum_{i=1}^{N}e^{j\alpha(\hat{t}_{N+1}-t_i)}g^*_{i}(t_1)g_{N+1}(t_1) +\sum_{i=1}^{N}e^{j\alpha(t_i-\hat{t}_{N+1})}g_{i}(t_1)g^*_{N+1}(t_1)
  \right]\\
  \nonumber &=&(j\alpha)^2\left[ 2\left(
  \sum_{i=1}^{N}e^{j\alpha t_i}g_i(t_1)g^*_{N+1}(t_1)
  \sum_{k=1}^{N}e^{-j\alpha t_{k}}g^*_{k}(t_1)
  g_{N+1}(t_1)\right)^{\frac{1}{2}}\right]\\
  \nonumber &=&-2\alpha^2\left|
  \sum_{i=1}^{N}e^{j\alpha t_i}g_i(t_1)g^*_{N+1}(t_1)\right|<0.
 \end{eqnarray}
 Therefore by substituting $\hat t_{N+1}$ and some simple manipulations we get
 \begin{align*}   \max_{t_{N+1}}q^2(\bt)=q^2(t_1,...,t_N,\hat{t}_{N+1})&=\left|\sum_{i=1}^{N}e^{j\alpha t_i}g_i(t_1)\right|^2+\left|g_{N+1}(t_1)\right|^2\\
   &+ 2\left(\sum_{i=1}^{N}e^{j\alpha t_i}g_i(t_1)g^*_{N+1}(t_1)
   \sum_{k=1}^{N}e^{-j\alpha t_{k}}g^*_{k}(t_1)
   g_{N+1}(t_1)\right)^{\frac{1}{2}}\\
   &=\left|\sum_{i=1}^{N}e^{j\alpha t_i}g_i(t_1)\right|^2+\left|g_{N+1}(t_1)\right|^2+2\left|\sum_{i=1}^{N}e^{j\alpha t_i}g_i(t_1)g^*_{N+1}(t_1)\right|\\
   &=\left|\sum_{i=1}^{N}e^{j\alpha t_i}g_i(t_1)\right|^2+\left|g_{N+1}(t_1)\right|^2+2\left|g_{N+1}(t_1)\right|\left|\sum_{i=1}^{N}e^{j\alpha t_i}g_i(t_1)\right|\\
   &=\left(\left|\sum_{i=1}^{N}e^{j\alpha t_i}g_i(t_1)\right|+\left|g_{N+1}(t_1)\right|\right)^{2}.
  \end{align*}
  As a result,
  \begin{eqnarray*}
   \max_{\left\{t_2,\dots,t_{N+1}\right\}}q(\bt)&=&\max_{\left\{t_2,\dots,t_N\right\}}\max_{t_{N+1}}q(\bt)=\max_{\left\{t_2,\dots,t_N\right\}} \left\{\left|\sum_{i=1}^{N}e^{j\alpha t_i}g_i(t_1)\right|\right\}+\left|g_{N+1}(t_1)\right|\\
   &=&\sum_{i=1}^{N}\left|g_i(t_1)\right|+\left|g_{N+1}(t_1)\right|=\sum_{i=1}^{N+1}\left|g_i(t_1)\right|.
  \end{eqnarray*}
\end{enumerate}

%
%
%
%
%
%

\section{Proof of Lemma \ref{lemma:PA_est}} \label{app:lemma:PAExtended}
\begin{enumerate}
  \item By recalling the definition of $S_n(\btau)=\sum_{m=1}^{N_t} \frac{1}{(c\tau_{m,n})^\beta}\hat s_m e^{j2\pi f_c \left(\tau_{1,1}-\tau_{m,n}\right)}$ and the pdf of $h$ given by $f_h(h)=\frac{1}{\sqrt{\pi}}e^{-|h|^2}$, the
likelihood under hypothesis $\mathcal{H}_1$ is given by
\begin{align*}
  f_1(\bR \med {\bold
  \btau},h)=(\pi)^{-KN_r}\exp\Bigg(-\underset{\dff D_2(\btau,h)}{\underbrace{\sum_{n=1}^{N_r}\sum_{k=1}^{K}\bigg|r_n[k]-\sqrt{\frac{E}{N_t}}S_n(\btau)h \;s[k;\tau_{1,1}]\bigg|^2}}\Bigg).
\end{align*}
We further simplify $D_3(\btau,h)$ as follows.
\begin{align}
  \nonumber D_2(\btau,h)
  &=\sum_{n=1}^{N_r}\sum_{k=1}^{K}\bigg|r_n[k]-\sqrt{\frac{E}{N_t}}S_n(\btau)h \;s[k;\tau_{1,1}]\bigg|^2\\
  \nonumber
  &=\sum_{k=1}^K\|\br[k]\|^2 + |h|^2 \bigg\{\frac{E}{T_sN_t}\sum_{n=1}^{N_r}| S_n(\btau)|^2\bigg\}\\
  \nonumber &-h^*\sqrt{\frac{E}{N_t}}\sum_{n=1}^{N_r}S^*_n(\btau) \sum_{k=1}^{K} r_n[k]s^*[k;\tau_{1,1}]
  -h\sqrt{\frac{E}{N_t}}\sum_{n=1}^{N_r}S_n(\btau) \sum_{k=1}^{K}
  r^*_n[k]s[k;\tau_{1,1}].
\end{align}
Therefore, the MAP estimate of $h$ is given by
\begin{align}
  \nonumber \hat h&=\arg\max_{h}\left\{ f_1(\bR\med {\boldsymbol{\tau}},h)f_{h}(h)\right\} =\arg\min_{h}\; \left\{\hat D_2(\btau,h)+|h|^2\right\}\\
  \label{eq:PAalpha}&=\frac{\sqrt{\frac{E}{N_t}}\sum_{n=1}^{N_r}S^*_n(\btau) \sum_{k=1}^{K} r_n[k]s^*[k;\tau_{1,1}]}
  {\frac{E}{T_s N_t}\sum_{n=1}^{N_r}| S_n(\btau)|^2+1},
\end{align}
which in turn after some manipulations provides that
\begin{eqnarray}
  \nonumber \hat D_2(\btau,\hat h)+|\hat h|^2 &=&
  \sum_{k=1}^{K}\|\br[k]\|^2-
  |\hat h|^2 \bigg\{\frac{E}{T_sN_t}\sum_{n=1}^{N_r}| S_n(\btau)|^2\bigg\}.
\end{eqnarray}
Therefore,
\begin{align}
  \nonumber \hat\btau^{\rm MAP}&=\arg\max_{\btau\in{\cal A'}({\cal X}')}\left\{ f_1(\bR\med {{\btau}},\hat h)f_{h}(\hat h)\right\} =\arg\max_{\btau\in{\cal A'({\cal X}')}} |\hat h|^2 \bigg\{\frac{E}{T_sN_t}\sum_{n=1}^{N_r}| S_n(\btau)|^2+1\bigg\}\\
  \label{eq:EstimatorPhasedArrayExtended}&=\arg\max_{\btau\in{\cal A}'({\cal X}')} \frac{\left|\sum_{n=1}^{N_r}S^*_n(\btau)\sum_{k=1}^{K} r_n[k]s^*[k;\tau_{1,1}]\right|^2}
  {\frac{1}{T_s}\sum_{n=1}^{N_r}|S_n(\btau)|^2+\frac{N_t}{E}},
\end{align}
which is the desired result.
  \item By recalling the definition of $S_n(\btau)=\sum_{m=1}^{N_t} \frac{1}{(c\tau_{m,n})^\beta}\hat s_m e^{j2\pi f_c \left(\tau_{1,1}-\tau_{m,n}\right)}$ and the pdf of $h$ given by $f_h(h)=\frac{1}{\sqrt{\pi}}e^{-|h|^2}$, the
likelihood under hypothesis $\mathcal{H}_1$ is given by
\begin{align*}
  f_1(\bR\med {\boldsymbol{\tau}})&=\int_h f_1(\bR \med {\bold
  \tau},h)f_h(h)\;dh\\
  &=\frac{(\pi)^{-KN_r}}{\sqrt\pi}\int_h
  \exp\Bigg(-\underset{\dff D_3(\btau)}{\underbrace{\sum_{n=1}^{N_r}\sum_{k=1}^{K}\bigg|r_n[k]-\sqrt{\frac{E}{N_t}}S_n(\btau)h \;s[k;\tau_{1,1}]\bigg|^2+|h|^2}}\Bigg)\;dh.
\end{align*}
We further simplify $D_3(\btau,\zeta)$ as follows.
\begin{align}
  \nonumber D_3(\btau,\zeta)
  &=\sum_{n=1}^{N_r}\sum_{k=1}^{K}\bigg|r_n[k]-\sqrt{\frac{E}{N_t}}S_n(\btau)h \;s[k;\tau_{1,1}]\bigg|^2+|h|^2\\
  \nonumber
  &=\sum_{k=1}^K\|\br[k]\|^2 + |h|^2 \bigg\{\frac{E}{T_sN_t}\sum_{n=1}^{N_r}| S_n(\btau)|^2\;+1\bigg\}\\
  \nonumber &-h^*\sqrt{\frac{E}{N_t}}\sum_{n=1}^{N_r}S^*_n(\btau) \sum_{k=1}^{K} r_n[k]s^*[k;\tau_{1,1}]
  -h\sqrt{\frac{E}{N_t}}\sum_{n=1}^{N_r}S_n(\btau) \sum_{k=1}^{K}
  r^*_n[k]s[k;\tau_{1,1}]\\
  \nonumber &=\sum_{k=1}^K\|\br[k]\|^2+\Bigg| h\sqrt{\frac{E}{T_s\cdot N_t}\sum_{n=1}^{N_r}|S_n(\btau)|^2+1}-\frac{\sqrt{\frac{E}{N_t}} \sum_{n=1}^{N_r}S^*_n(\btau) \sum_{k=1}^{K} r_n[k]s^*[k;\tau_{1,1}]}{\sqrt{\frac{E}{T_s\cdot N_t}\sum_{n=1}^{N_r}|  S_n(\btau)|^2+1}}\Bigg|^2\\
  \label{eq:exponentLPAExtended} &-\frac{\frac{E}{N_t} \left|\sum_{n=1}^{N_r}S^*_n(\btau) \sum_{k=1}^{K} r_n[k]s^*[k;\tau_{1,1}]\right|^2}{\frac{E}{T_s\cdot N_t}\sum_{n=1}^{N_r}|S_n(\btau)|^2+1}.
\end{align}

By following the same line of argument as in the proof of Theorem
\ref{th:multi_ML} it can be readily demonstrated that the ML
estimate of the time-delay for phased-array system is given by
\begin{eqnarray*}
  \hat\btau^{\rm ave}&=&\arg\max_{\btau\in{\cal A}'({\cal X}')}f_1(\bR\med {\boldsymbol{\tau}})\\
  &=&\arg\max_{\btau\in{\cal A}'({\cal X}')}\left\{\frac{ \left|\sum_{n=1}^{N_r} S^*_n(\btau) \sum_{k=1}^{K} r_n[k]s^*[k;\tau_{1,1}]\right|^2}{\frac{1}{T_s}\sum_{N=1}^{N_r}|S_n(\btau)|^2+\frac{N_t}{E}}-\frac{1}{2}\log\left(\frac{E}{T_s N_t}\sum_{n=1}^{N_r}|
  S_n(\btau)|^2+1\right)\right\}.
\end{eqnarray*}
  \item Setting $P=1$ in (\ref{eq:model_cont}), it is seen that
$h_{m,n}=\zeta\;e^{-j2\pi f_c\tau_{m,n}}$. Therefore, the likelihood
under hypothesis $\mathcal{H}_1$ becomes
\begin{align*}
  f_1(\bR \med {\bold
  \btau},\zeta)=(\pi)^{-KN_r}
  \exp\Bigg(-\underset{\dff D_4(\btau,\zeta)}{\underbrace{\sum_{n=1}^{N_r}\sum_{k=1}^{K}\bigg|r_n[k]-\zeta\sqrt{\frac{E}{N_t}}s[k;\tau_{1,1}]\sum_{m=1}^{N_t}\frac{1}{(c\tau_{m,n})^\beta} \hat s_m e^{j2\pi f_c (\tau_{1,1}-2\tau_{m,n})}\bigg|^2}}\Bigg).
\end{align*}
By defining $\hat S_n(\btau)\dff \sum_{m=1}^{N_t}
\frac{1}{(c\tau_{m,n})^\beta}\hat s_m e^{j2\pi f_c
(\tau_{1,1}-2\tau_{m,n})}$ and following the same line of argument as in the previous part the ML estimate for $\zeta$ is found as
\begin{align}
  \nonumber\hat\zeta&=\arg\max_{\zeta}\left\{\log f_1(\bR\med {\boldsymbol{\tau}},\zeta)\right\} =\arg\min_{\zeta}\; \left\{D_4(\btau,\zeta)\right\}=\frac{\sum_{n=1}^{N_r}\hat S^*_n(\btau) \sum_{k=1}^{K} r_n[k]s^*[k;\tau_{1,1}]}
  {\frac{1}{T_s}\sqrt{\frac{E}{N_t}}\sum_{n=1}^{N_r}|\hat S_n(\btau)|^2},
\end{align}
and consequently
\begin{eqnarray}
  \nonumber D_4(\btau,\hat\zeta) &=&
  \sum_{k=1}^{K}\|\br[k]\|^2-
  \frac{E}{T_s N_t}|\hat{\zeta}|^2\sum_{n=1}^{N_r}|\hat S_n(\btau)|^2.
\end{eqnarray}
Therefore by following a similar approach as in part 1 we get
\begin{align}
  \nonumber \hat\btau&=\arg\max_{\btau\in{\cal A}'({\cal X}')}f_1(\bR\med {{\btau}},\hat\zeta) =\arg\max_{\btau\in{\cal A}'} \bigg\{\frac{E}{T_s N_t}|\hat\zeta|^2\sum_{n=1}^{N_r}|\hat S_n(\btau)|^2\bigg\}\\
  \label{eq:EstimatorPhasedArrayExtended}&= \arg\max_{\btau\in{\cal A}'({\cal X}')} \frac{\left|\sum_{n=1}^{N_r}\hat S^*_n(\btau)\sum_{k=1}^{K} r_n[k]s^*[k;\tau_{1,1}]\right|^2}
  {\frac{1}{T_s}\sum_{n=1}^{N_r}|\hat S_n(\btau)|^2},
\end{align}
which is the desired result.
\end{enumerate}

\section{Proof of Lemma \ref{lemma:diversity}}
\label{app:lemma:diversity}

We define the events
\begin{eqnarray*}
   \mathcal{D}_1\dff \bigcup_{m=1}^M\left\{Y_m^2<\gamma\right\},\;\;\;
   \mbox{and}\;\;\;\mathcal{D}_2\dff \left\{\sum_{m=1}^MY_m^2<\gamma\right\},\;\;\;
   \mbox{and}\;\;\;\mathcal{D}_3\dff \bigcup_{m=1}^M\left\{Y_m^2<\frac{\gamma}{M}\right\}.
\end{eqnarray*}
Clearly as $\mathcal{D}_3\subseteq\mathcal{D}_2\subseteq\mathcal{D}_1$ we have $\pr(\mathcal{D}_3)\leq \pr(\mathcal{D}_2)\leq \pr(\mathcal{D}_1)$, where $\pr(\mathcal{D}_2)$ is the probability of interest in (\ref{eq:diversity}). In order to analyze $P(\mathcal{D}_1)$ we define random variables $\hat Y_m\sim\mathcal{N}(\rho\cdot|\mu_m|,\sigma^2)$, for which we have
\begin{eqnarray}
    \nonumber \pr(Y_m^2<\gamma)&=&\pr(\hat Y_m^2<\gamma)=\pr(-\sqrt{\gamma}<\hat Y_m<\sqrt{\gamma})= Q\left(\frac{\rho|\mu_m|-\sqrt\gamma}{\sigma}\right)-Q\left(\frac{\rho|\mu_m|+\sqrt\gamma}{\sigma}\right)\\
    \label{eq:diversity:proof1}& \circeq & Q\left(\frac{\rho|\mu_m|-\sqrt\gamma}{\sigma}\right)\\
    \label{eq:diversity:proof2}& \circeq & Q\left(\frac{\rho|\mu_m|}{\sigma}\right),
\end{eqnarray}
where the transition in (\ref{eq:diversity:proof1}) holds by noting that for any choice of $\gamma>\sigma^2$, from the table of $Q(\cdot)$ function it is readily verified that for sufficiently large values of $x$, and any fixed value $\Delta>0$, $Q(x)\gg Q(x-\Delta)$. The asymptotic equality in (\ref{eq:diversity:proof2}) is justified by noting that $\rho|\mu_m|\gg \sqrt\gamma$. By further using the following known bounds on the $Q(\cdot)$
\begin{equation*}
    \forall x\in\mathbb{R}^+,\quad \frac{1}{\sqrt{2\pi}x}\underset{\circeq 1}{\underbrace{\left(1-\frac{1}{x^2}\right)}}e^{-x^2/2}\leq Q(x) \leq \frac{1}{\sqrt{2\pi}x}e^{-x^2/2},
\end{equation*}
we get
\begin{equation*}
    \forall x\in\mathbb{R}^+,\quad \frac{1}{\sqrt{2\pi}x}e^{-x^2/2}\;\circlt\; Q(x)\; \circlt\; \frac{1}{\sqrt{2\pi}x}e^{-x^2/2}\quad\Rightarrow\quad Q(x)\circeq \frac{1}{\sqrt{2\pi}x}e^{-x^2/2}.
\end{equation*}
By further taking into account that $\frac{1}{\sqrt{2\pi}x}e^{-x^2/2}\doteq e^{-x^2/2}$ and noting Remark \ref{remark:1} we get
\begin{equation}\label{eq:Q_bounds}
    Q(x)\doteq e^{-x^2/2}.
\end{equation}

Invoking the exponential equivalence in (\ref{eq:Q_bounds}) on (\ref{eq:diversity:proof2}) provides \begin{equation}\label{eq:diversity:proof3}
    \pr(Y_m^2<\gamma)\doteq e^{-\frac{\rho^2\mu_m^2}{2\sigma^2}}.
\end{equation}
As $\mu_m$ is an unknown random variable distributed as $\mathcal{N}(0,\sigma^2_\mu)$, by averaging out its effect we get
\begin{equation}\label{eq:diversity:proof4}
    \bbe_{\mu_m}\left[\pr(Y_m^2<\gamma)\right]\doteq \frac{1}{\sqrt{2\pi}}\int_{-\infty}^\infty e^{-(\frac{\rho^2}{\sigma^2}+\frac{1}{\sigma^2_m})\frac{\mu_m^2}{2}}d\mu_m=\frac{1}{\sqrt{\frac{\rho^2}{\sigma^2}+\frac{1}{\sigma^2_m}}}\doteq \rho^{-1}.
\end{equation}
Next, by considering the statistical independence of $\{Y_m\}_{m=1}^M$ we consequently find
\begin{equation}\label{eq:diversity:proof5}
    \bbe_{\bmu}\left[\pr(\mathcal{D}_1)\right]=\prod_{m=1}^M\bbe_{\mu_m}\left[\pr(Y_m^2<\gamma)\right]\doteq
    \rho^{-M}.
\end{equation}

Note that, as long as the asymptotic behavior is concerned, the
asymptotic exponential order of
$\bbe_{\bmu}\left[\pr(\mathcal{D}_1)\right]$ does not depend on the
choice of $\gamma$ and by following the same line of argument as
above we can extend the same result for the choice of
$\frac{\gamma}{M}$ and thereof for the exponential order of
$\bbe_{\bmu}\left[\pr(\mathcal{D}_3)\right]$. As a result, we can
similarly show that
\begin{equation}\label{eq:diversity:proof6}
    \bbe_{\bmu}\left[\pr(\mathcal{D}_3)\right]\doteq \rho^{-M}.
\end{equation}

By noting that $\pr(\mathcal{D}_3)\leq \pr(\mathcal{D}_2)\leq \pr(\mathcal{D}_1)$ and using (\ref{eq:diversity:proof5}) and (\ref{eq:diversity:proof6}) we get that
\begin{equation*}
    \pr(\mathcal{D}_3)\;\dotlt\; \pr(\mathcal{D}_2)\;\dotlt \; \pr(\mathcal{D}_1)\quad\Rightarrow\quad \pr\left(\sum_{m=1}^MY_m^2<\gamma\right)=\pr(\mathcal{D}_2)\doteq\rho^{-M},
\end{equation*}
which concludes the proof.

\renewcommand\url{\begingroup\urlstyle{rm}\Url}
\renewcommand{\baselinestretch}{1.4}
{\small \bibliographystyle{IEEEtran}
\bibliography{IEEEabrv,MIMO_Radar}}

\begin{thebibliography}{10}
\providecommand{\url}[1]{#1}
\csname url@samestyle\endcsname
\providecommand{\newblock}{\relax}
\providecommand{\bibinfo}[2]{#2}
\providecommand{\BIBentrySTDinterwordspacing}{\spaceskip=0pt\relax}
\providecommand{\BIBentryALTinterwordstretchfactor}{4}
\providecommand{\BIBentryALTinterwordspacing}{\spaceskip=\fontdimen2\font plus
\BIBentryALTinterwordstretchfactor\fontdimen3\font minus
  \fontdimen4\font\relax}
\providecommand{\BIBforeignlanguage}[2]{{%
\expandafter\ifx\csname l@#1\endcsname\relax
\typeout{** WARNING: IEEEtran.bst: No hyphenation pattern has been}%
\typeout{** loaded for the language `#1'. Using the pattern for}%
\typeout{** the default language instead.}%
\else
\language=\csname l@#1\endcsname
\fi
#2}}
\providecommand{\BIBdecl}{\relax}
\BIBdecl

\bibitem{Fishler:RC04}
E.~Fishler, A.~Haimovich, R.~Blum, L.~Cimini, D.~Chizhik, and R.~Valenzuela,
  ``{MIMO} radar: {A}n idea whode time has come,'' in \emph{Proc. of IEEE Radar
  Conference}, Philadelphia, PA, April 2004.

\bibitem{Fishler:Asilomar04}
E.~Fishler, A.~M. Haimovich, R.~Blum, L.~Cimini, D.~Chizhik, and R.~Valenzuela,
  ``Performance of {MIMO} radar systems: {A}dvantes of angular diversity,'' in
  \emph{Proc. 38th Asilomar Conf. on Signals, Systems and Computers}, Pacific
  Grove, CA, November 2004.

\bibitem{Fishler:SP06}
E.~Fishler, A.~Haimovich, and R.~S. Blum, ``Spatial diversity in radars--models
  and detection performance,'' \emph{{IEEE} Trans. Signal Process.}, vol.~54,
  no.~3, pp. 823--837, March 2006.

\bibitem{Haimovich:SPM08}
A.~M. Haimovich, R.~S. Blum, and L.~J. Cimini, ``{MIMO} radar with widely
  seperated antennas,'' \emph{{IEEE} Signal Process. Mag.}, vol.~25, no.~1, pp.
  116 -- 129, January 2008.

\bibitem{Lehmann:Asilomar06}
N.~H. Lehmann, A.~M. Haimovich, R.~S. Blum, and L.~Cimini, ``High resolution
  capabilities of {MIMO} radar,'' in \emph{Proc. 38th Asilomar Conf. on
  Signals, Systems and Computers}, Pacific Grove, CA, November 2006.

\bibitem{Stoica:SP07}
P.~Stoica, J.~Li, and Y.~Xie, ``On probing signal design for {MIMO} radar,''
  \emph{{IEEE} Trans. Signal Process.}, vol.~55, no.~8, pp. 4151 -- 4161,
  August 2007.

\bibitem{Li:SPL07}
J.~Li, P.~Stoica, L.~Xu, and W.~Roberts, ``On parameter identifiability of
  {MIMO} radar,'' \emph{{IEEE} Signal Process. Lett.}, vol.~14, no.~12, pp. 968
  -- 971, December 2007.

\bibitem{Godrich:IT08_submitted}
H.~Godrich, A.~M. Haimovich, and R.~S. Blum, ``Target localization accuracy
  gain in {MIMO} radar based systems,'' \emph{{IEEE} Trans. Inf. Theory}, 2008,
  submitted.

\bibitem{Li:SPM07}
J.~Li and P.~Stoica, ``{MIMO} radar with collocated antennas,'' \emph{{IEEE}
  Signal Process. Mag.}, vol.~24, no.~5, pp. 106--114, September 2007.

\bibitem{Bekkerman:SP06}
I.~Bekkerman and J.~Tabrikian, ``Target detection and localization using {MIMO}
  radars and sonars,'' \emph{{IEEE} Trans. Signal Process.}, vol.~54, no.~10,
  pp. 3873--3883, 2006.

\bibitem{Levy:book}
B.~C. Levy, \emph{Principles of Signal Detection and Parameter Estimation},
  1st~ed.\hskip 1em plus 0.5em minus 0.4em\relax Springer, 2008.

\bibitem{Skolnik:book}
M.~Skolnik, \emph{Introduction to {R}adar {S}ystems}, 3rd~ed.\hskip 1em plus
  0.5em minus 0.4em\relax McGraw-Hill, 2002.

\bibitem{Haykin:book}
S.~Haykin, J.~Litva, and T.~J. Shepherd, \emph{Radar {A}rray {P}rocessing},
  1st~ed.\hskip 1em plus 0.5em minus 0.4em\relax New York: Springer-Verilog,
  1993.

\bibitem{Moustakides:IT09_submitted}
G.~V. Moustakdies, ``Finite sample size optimality of {GLR} tests,''
  \emph{{IEEE} Trans. Inf. Theory}, 2009, submitted. {Available online at
  \url{http://arxiv.org/abs/0903.3795}}.

\bibitem{Kendall:book}
M.~Kendall, A.~Stuart, and S.~Arnold, \emph{Advanced Theory of Statistics,
  Classical Inference and the Linear Model}.\hskip 1em plus 0.5em minus
  0.4em\relax New York: Hodder Arnold Publications, 1999, vol.~2A.

\bibitem{Poor:book}
H.~V. Poor, \emph{An Introduction to Signal Detection and Estimation},
  2nd~ed.\hskip 1em plus 0.5em minus 0.4em\relax Springer-Verlag, 1998.

\bibitem{Balazs:online}
M.~Balazs, ``Sum of independent exponential random variables with different
  parameters,'' available at \url{http://www.math.bme.hu/~balazs/sumexp.pdf}.

\bibitem{Svecova:Radio08}
M.~Svecova, K.~D, and R.~Zetik, ``Object localization using round trip
  propagation time measurements,'' in \emph{Proc. Radioelektronika, 18th
  International Conference}, Czech Republic, April 2008, pp. 1--4.

\end{thebibliography}

\end{document}